\documentclass{aa} 
\usepackage{float} 
\usepackage{graphicx}  
\usepackage{graphics}
\usepackage{txfonts}  
\usepackage{natbib}
\usepackage{adjustbox}
\usepackage{booktabs}
\usepackage{mathtools}
\usepackage{comment}
\bibpunct{(}{)}{;}{a}{}{,} 
\usepackage{multicol}
\usepackage{placeins}
\usepackage{longtable}
\usepackage[dvipsnames]{xcolor}
\usepackage[normalem]{ulem}
\usepackage{bm}
\usepackage{braket}
\usepackage{tabularx}
\usepackage{hyperref}
\hypersetup{
    colorlinks=true,
    linkcolor=blue,
    citecolor=blue,
    filecolor=magenta,      
    urlcolor=cyan,
    pdftitle={Overleaf Example},
    pdfpagemode=FullScreen,
    }

\begin{document}

\title {Glitches in solar-like oscillating F-type stars}
\subtitle{Possible contribution of non-linear terms}

\author{M. Deal\inst{1} \and M.-J. Goupil\inst{2} \and J. Philidet\inst{2} \and M. S. Cunha\inst{3} \and R. Teissonniere\inst{1}\and E. Josselin\inst{1}}
  
\institute{LUPM, Universit\'e de Montpellier, CNRS, Place Eug\`ene Bataillon, 34095 Montpellier, France\\
\email{morgan.deal@umontpellier.fr} 
           \and
LESIA, Observatoire de Paris, Universit\'e PSL, CNRS, Sorbonne Universit\'e, Universit\'e de Paris, 5 place Jules Janssen, 92195 Meudon, France
           \and
Instituto de Astrofísica e Ciências do Espaço, Universidade do Porto,
CAUP, Rua das Estrelas, PT4150-762 Porto, Portugal 
            }
           
\date{\today}

\abstract
{The glitch signatures in $r_{010}$ for F-type stars (higher amplitude and period of the oscillatory component) are very different from those of G-type stars.}
{The aim of this work is to analyse the signatures of these glitches and understand the origin of the differences in these signatures between G-type and F-type stars.}
{We fit the glitch signatures in the frequencies, second differences, and $r_{010}$ ratios while assuming either a sinusoidal variation or a more complex  expression. The fit provides the acoustic depth, and hence the position, of the bottom of the convective envelope for nine \textit{Kepler} stars and the Sun.}
{We find that for F-type stars, the most commonly used fitting expressions for the glitch of the bottom of the convective envelope provide different measurements of the position of the bottom of the convective envelope for the three seismic indicators, while it is not the case for G-type stars. When adding an additional term in the fitting expression with twice the acoustic depth of the standard term (a contribution that accounts for the highly non-sinusoidal shape of the signature in the $r_{010}$ ratios), we find better agreement between the three seismic indicators and with the prediction of stellar evolution models.}
{While the origin of this additional term is not yet understood, this may be an indication that the transition between the convective envelope and the underlying radiative zone is different for G- and F-type stars. This outcome brings new insight into the physics in these regions.}

  
  

\maketitle 

\nolinenumbers
\section{Introduction}
Stochastically excited pressure modes provide a wealth of information about the internal structure of solar-like stars. Furthermore, variations of the internal structure of stars on a scale smaller than the wavelength of the modes, the so-called glitches, induce perturbations of the oscillation frequencies that can be detected and measured. Analysis of this signature enables a direct measurement of the position of the glitch inside the star and yields information about the internal structure in this region.

Studying the glitch signatures is particularly powerful for inferring the helium content in the envelope, thanks to the glitch of the second ionisation zone in the first adiabatic exponent $\Gamma_1$ \citep[][and references therein]{gough90,verma17,verma19}. The position of the convective boundaries can also be measured using glitches induced by the variation of the temperature and composition gradients \citep[][and references therein]{monteiro94, roxburgh94,roxburgh09,mazumdar14,deheuvels16,verma17,deal23}.

In \cite{deal23} (hereafter Paper I), the authors show that the signature of the base of the convective envelope for F-type stars differs from that for G-type stars (see their Fig.~1). While the signature in G-type stars shows a low-amplitude and a quasi-sinusoidal shape most of the time, the signature in F-type stars corresponds to high-amplitude and non-sinusoidal variations. In addition, the authors deduce the depth of the base of the convective zone (BSCZ) from the glitch signature for G-type and F-type stars in the same way (i.e. ignoring the non-sinusoidal aspect in the latter case). The result shows that the BSCZ in F-type stars is much deeper (more than 1 to 2 $H_p$, with $H_p$ being the height of the pressure scale) than predicted by the standard models. This discrepancy can hardly be explained by assuming that the depth of the convective boundary is extended downwards by penetrative convection. Indeed, in such a case, one expects an extension of only about $0.3~H_p$ from 3D hydrodynamical simulations for F-type stars \citep[e.g.][]{breton22}. In fact, the non-sinusoidal shape of the signature of F-type stars gives rise to a multi-peak distribution of the position of the glitch in a Fourier spectrum, which makes it difficult to analyse. Non-sinusoidal variations may arise when adding components to the model that oscillate at harmonic frequencies of the basic frequency (not to be confused with overtones). This is the case, for example, when modelling intrinsically non-linear pulsations such as those of Cepheid or RR Lyrae stars, high amplitude Delta Scuti stars, or pulsating white dwarfs (for more details, see the reviews by \citealt{dziembowski93} or \citealt{smolec11}).

In this paper, we explore the consequences of going beyond a linear approach by adding the contribution of the first order term beyond linearity to the theoretical expression representing the glitch signature in the variations of the oscillation mode frequency with radial order, in the variations of the second difference, and of the $r_{010}$ ratios with frequency. Assuming the signatures are linked to the bottom of the convective envelope, the idea is to determine if this term results in a decrease of the measured extension of the convective penetration and leads to a measured value closer to the theoretical expectations and predictions from 3D simulations.

In Sect.~\ref{sect:theory}, we briefly describe the current theory of glitches induced by the variation of the temperature gradient at the boundary of convective zones. We emphasise the issues raised by the fit of such a signature in F-type stars in Sect.~\ref{sec:issue}. We define the targets and the expression of the non-sinusoidal terms that we think are needed to understand the glitch signature in F-type stars in Sect.~\ref{sec:sampleA} and \ref{sec:sampleB}. We then turn to observational data and compare the measurements resulting from the fits to predictions from stellar models in Sect.~\ref{sect:comp_obs}. We finally discuss some possible origins for these non-sinusoidal terms in Sect.~\ref{sect:discu} and conclude in Sect.~\ref{sect:conclu}.

\section{Position of the base of the convective zone}\label{sect:theory}
The measurement of the position of BSCZ can be performed using the signature of the rapid variation of the sound speed (and its derivatives) occuring at the transition between the convective and the underlying radiative regions. At this transition, both the temperature and composition gradient vary over a region narrower than the wavelength of the modes. This induces a signature in the mode frequencies, in the second differences and, in the $r_{010}$ ratios. Assuming a step function 
for the transition from the adiabatic temperature gradient in the convective envelope to the radiative ones, the signature for a given mode frequency can be modelled by \citep{monteiro94,roxburgh94}
\begin{equation}\label{eq:mario}
    \begin{aligned}
    \delta\nu_\mathrm{cz}(\nu)\approx~~&a_1(\tau_\mathrm{cz})\left(\frac{\tilde{\nu}}{\nu}\right)^2 ~\sin \Phi_{cz} 
    +~~ a_2(\tau_\mathrm{cz})\left(\frac{\tilde{\nu}}{\nu}\right) ~\cos \Phi_{cz},
    \end{aligned}    
\end{equation}
\noindent
where $$\Phi_{cz} = 4\pi\nu\tau_\mathrm{cz}+2\phi$$
and $\tilde{\nu}$ is a reference frequency and $\phi$ some constant phase. $\tau_\mathrm{cz}$ is the acoustic depth of the BSCZ defined as
\begin{equation}\label{eq:tau}
    \tau_\mathrm{cz}=\int^{R}_{r_\mathrm{cz}}\frac{dr}{c_s} ~~;~\mathcal{T}=\int^{R}_{0}\frac{dr}{c_s}\approx1/(2\Delta\nu)
\end{equation}
\noindent where $r_\mathrm{cz}$ is the radius of the BSCZ and $R$ is the radius of the star to the acoustic surface (not to the photosphere),  $c_s$ is the sound speed and  $\mathcal{T}$  represents the total acoustic radius of the star. The constant amplitudes $a_1$ and $a_2$ are related to the physical conditions at the transition and the expressions can be found in \cite{monteiro94}, \cite{deal23}, and references therein. 

\subsection{The frequency ratios}

The scaled small separation ratios $r_{01}$ and $r_{10}$ (hereafter noted $r_{010}$) of a given mode with frequency $\nu$, angular degree $l=0$ and $1$, and radial order $n$ are defined as \citep{roxburgh03}:
\begin{equation}\label{eq:r010}
    r_{01}(n)=\frac{d_{01}(n)}{\Delta_1(n)}~~\mathrm{and}~~r_{10}(n)=\frac{d_{10}(n)}{\Delta_0(n+1)},
\end{equation}
\noindent where $\Delta_l(n)=\nu_{n,l}-\nu_{n-1,l}$ is the large separation and 
\begin{eqnarray}
    d_{01}(n)&=&\frac{1}{8}(\nu_{n-1,0}-4\nu_{n-1,1}+6\nu_{n,0}-4\nu_{n,1}+\nu_{n+1,0}), \\
    d_{10}(n)&=&-\frac{1}{8}(\nu_{n-1,1}-4\nu_{n,0}+6\nu_{n,1}-4\nu_{n+1,0}+\nu_{n+1,1}).
\end{eqnarray}
The glitch signature of the convective envelope is also detectable in these ratios and can be expressed as follows (Paper I):
\begin{equation}\label{eq:signal_r010}
\begin{aligned}
     r_{010,\mathrm{cz}}(\nu)=& a_1(\tau_\mathrm{cz})~~\Bigl(\frac{\tilde{\nu}}{\nu}\Bigr)^2 ~
    ~ \times\frac{f_{12}(\nu)}{4\Delta\nu} \times \sin \Phi_{cz} \\
        &+a_2(\tau_\mathrm{cz}) ~~\frac{\tilde{\nu}}{\nu}~ ~\times 
        \frac{f_{21}(\nu)}{4\Delta\nu} \times \cos \Phi_{cz} ,
\end{aligned}
\end{equation}
\noindent where $\Delta\nu$ is the mean large frequency separation and the expression of the frequency-dependent functions $f_{12}$ and $f_{21}$ can be found in Paper I.

\subsection{Second differences}

The second difference of a pressure mode with frequency $\nu$, angular degree $l$, and radial order $n$ is defined as \citep{gough90}:
\begin{equation}~\label{eq:sec_dif}
\Delta^2\nu_{n,\ell}  \vcentcolon  = \nu_{n{+}1,\ell} - 2 \nu_{n,\ell} + \nu_{n{-}1,\ell} \, .
\end{equation}
The glitch signature defined in Eq.~\ref{eq:mario} is also detectable in the second differences. The expression of the glitch signature in the second differences is defined as (Paper I):
\begin{equation}\label{eq:signal_sd}
  \begin{aligned}
  \Delta^2\nu_{n,l,cz}\approx~~& a_1(\tau_\mathrm{cz})\left(\frac{\tilde{\nu}}{\nu}\right)^2 \mathrm{ff}_{12}(\nu)\sin\left(\Phi_{cz}\right)+a_2(\tau_\mathrm{cz})\left(\frac{\tilde{\nu}}{\nu}\right) \mathrm{ff}_{21}(\nu)\cos\left(\Phi_{cz}\right)
  \end{aligned}    
\end{equation}
\noindent where the expression of the frequency depends on the functions $\mathrm{ff}_{12}$ and $\mathrm{ff}_{21}$, that can be found in Paper I. 

The second differences and the frequencies also include a signature of the glitch due to the helium second ionisation. For both of them, the helium signature dominates the signal and the one of the convective envelope is secondary. In contrast, for the $r_{010}$ calculated with five frequencies (as presented in the previous subsection), the signature of the helium ionisation zone is negligible and the one of the convective envelope dominates. In the following, when glitch signatures are fitted, the helium contribution is included for the second differences and the frequencies. The expressions are given in the following sections.

\begin{figure*}
    \centering
    \includegraphics[scale=0.61]{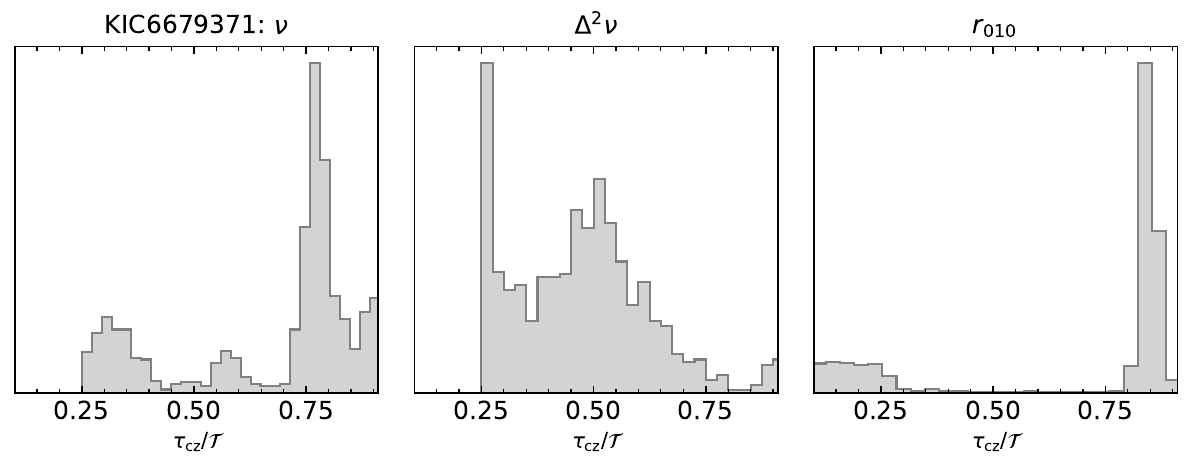}
    \includegraphics[scale=0.61]{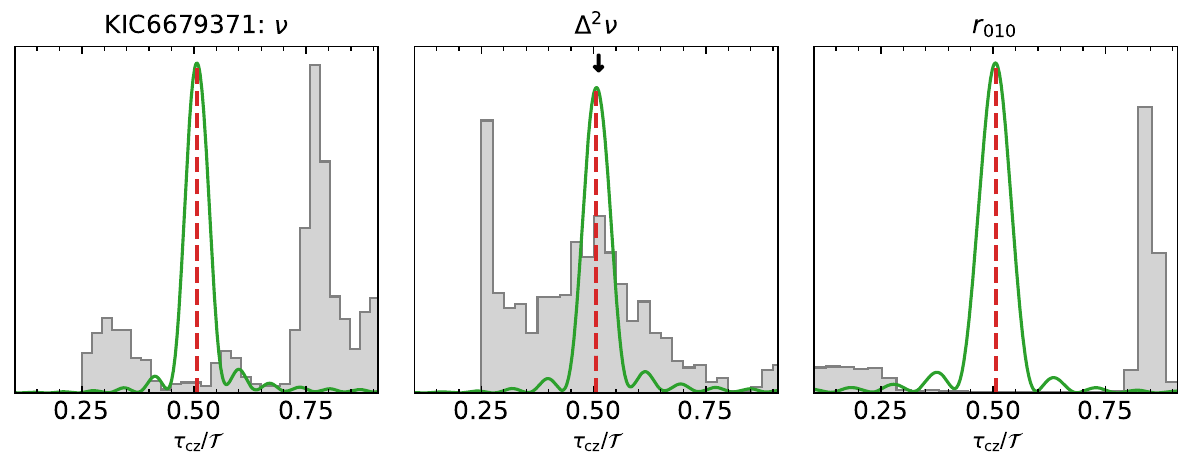}
    \includegraphics[scale=0.61]{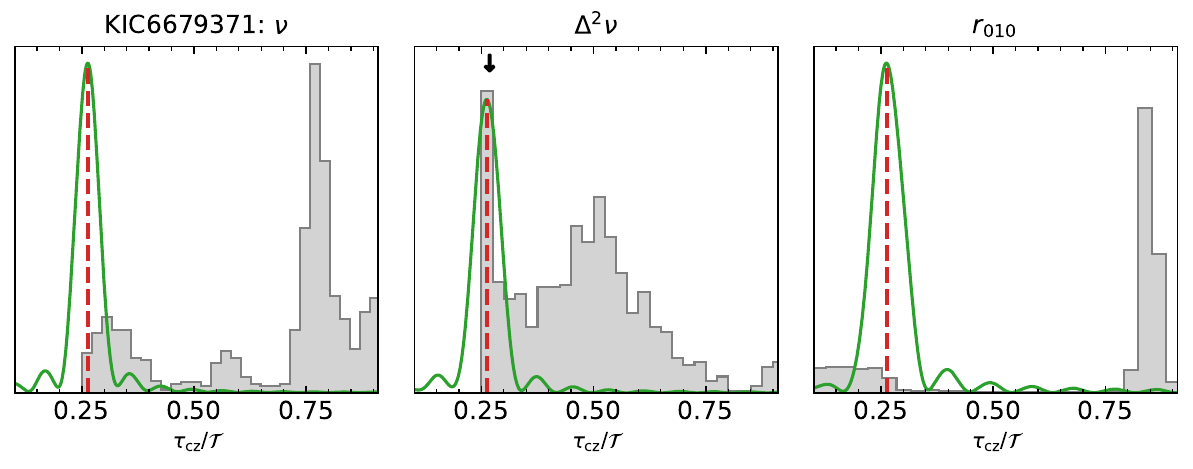}
    \includegraphics[scale=0.61]{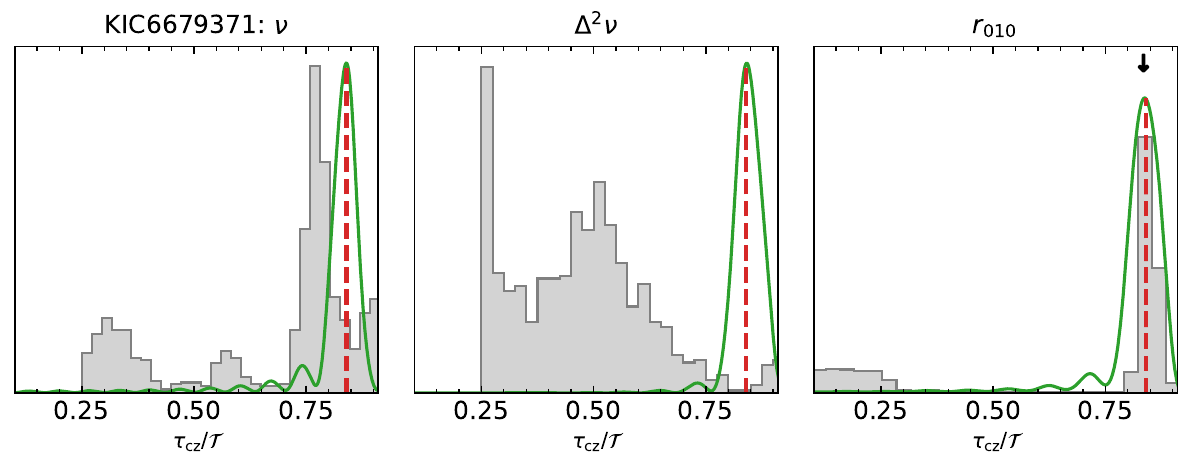}
   \caption{Distributions of $\tau_\mathrm{cz}/\mathcal{T}$ for the glitch signature in the frequencies (left panels), second differences (middle panels), and $r_{010}$ ratios (right panels) for the \textit{Kepler} star KIC6679371 (in light grey). $\mathcal{T}=1/(2\Delta\nu)$ is the total acoustic radius of the star. The results for the $r_{010}$ ratios are converted from the measured $t_\mathrm{cz}$ with $\tau_\mathrm{cz}$= $\mathcal{T}-t_\mathrm{cz}$. The light grey histograms are obtained by a fit of the observed data with the standard expressions (Eqs.~\ref{eq:fit_r010_std}, \ref{eq:fit_freq_std}, \ref{eq:fit_sd_std}). For clarity, the top panels represent the observed distributions alone. The green curve represent the resampled Fourier transform of the synthetic data. The red dashed vertical lines indicate the position of the BSCZ as inputted in the synthetic data. The black down arrows show the peak in the histogram of the observed distribution selected to generate the synthetic data.}
   \label{fig:tests_KIC667}
\end{figure*}

\begin{figure*}[ht]
    \centering
    \includegraphics[scale=0.625]{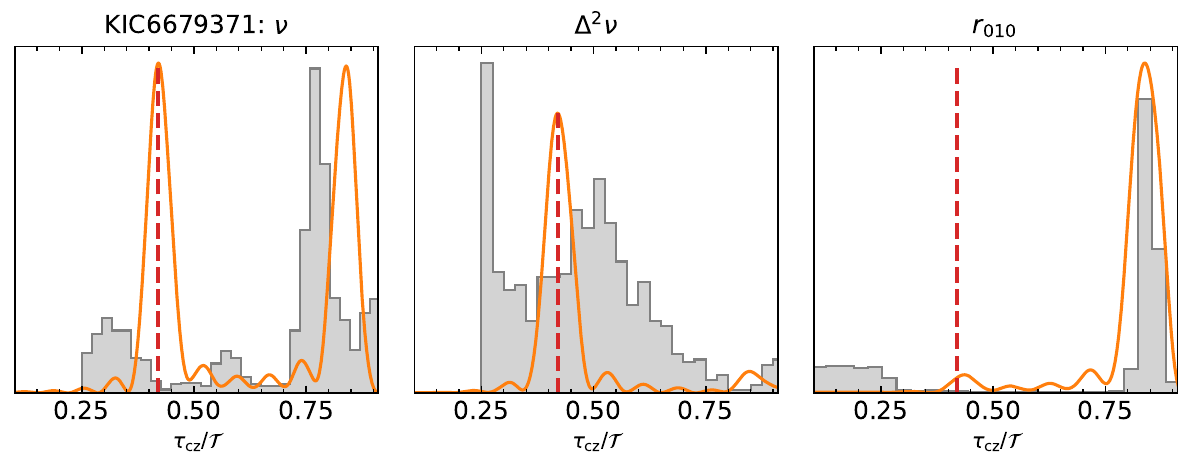}
    \includegraphics[scale=0.625]{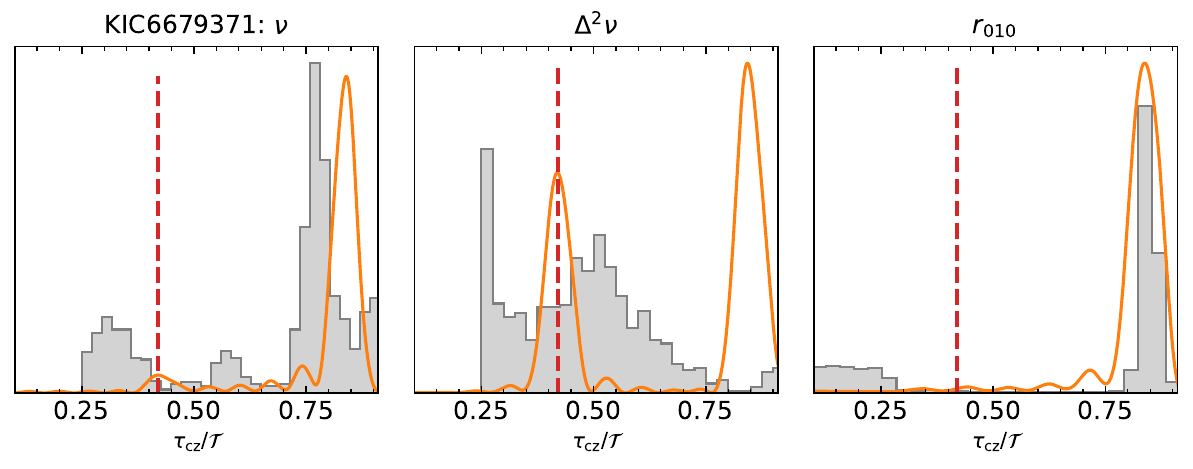}
    \includegraphics[scale=0.625]{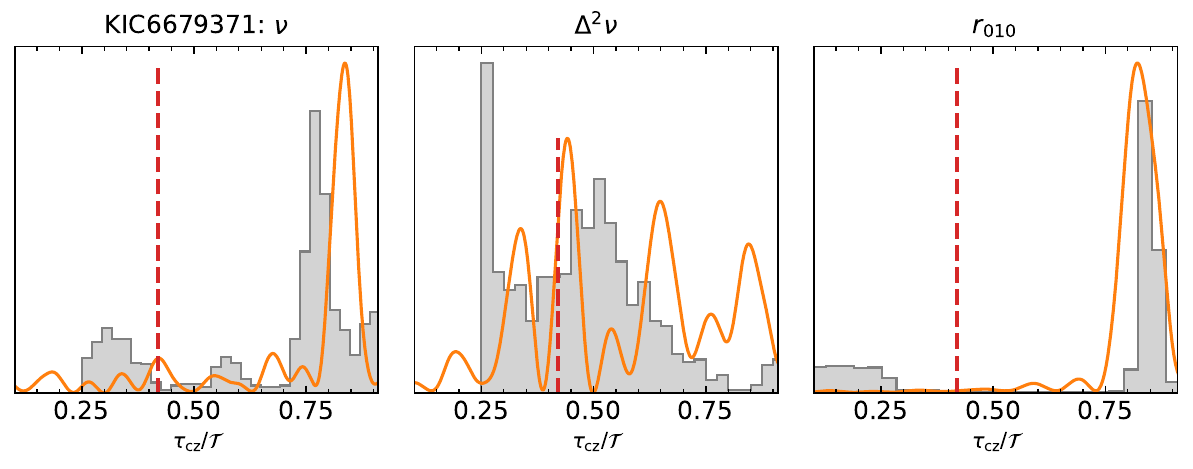}
   \caption{Same as Fig.~\ref{fig:tests_KIC667} except that the synthetic data include an additional term with twice the acoustic depth. The amplification factor $k=1$ and $5$ for the top and middle panel, respectively. For the bottom panel, $k=3$ and  random  noise is added to the synthetic data. The red dashed vertical lines indicate the position of the BSCZ as inputted in the synthetic data. That position is selected to reproduce the peaks in the distributions obtained from the frequencies and ratios with the additional term. The amplitude of the Fourier transform is adapted for clarity, especially for the second differences presented in the bottom panel for which the amplitude of the Fourier transform is smaller than for the other panels.}
   \label{fig:tests_KIC667_ns}
\end{figure*}

\section{Setting the problem}\label{sec:issue}

For the Sun and simple G-type stars (e.g. KIC8006161/Doris), the three indicators give the same results when they are fitted with the standard glitch expressions \citep[see][for the Sun]{monteiro94,roxburgh09}. The picture is different for F-type stars and we illustrate it with KIC66679371 in this section.
 
We first fit the observed data for KIC66679371 with the standard expression for the glitch signature and compare with the result of a Fourier transform  of a set of synthetic frequencies appropriately built. In a second step, we consider the case of a more complex formulation for the glitch signature.

\subsection{Fit of the observed data} 

\paragraph{Assuming a standard signature of the glitch}  
The fits of the observed data were performed using the standard expressions for the glitch signatures (hereafter called `standard fit'). 
The $r_{010}$ ratios are fitted with the procedure described in Paper I. The fitting expression is defined as  
\begin{equation}\label{eq:fit_r010_std}
\begin{aligned}
    r_{010}(\nu)=& P(\nu) + a_1(\tau_\mathrm{cz})~\Bigl(\frac{\tilde{\nu}}{\nu}\Bigr)^2
    ~ \times\frac{1}{4\Delta\nu} f_{12}(\nu)\times\sin\left(4\pi\nu t_\mathrm{cz}+2\phi\right)\\
        &+a_2(\tau_\mathrm{cz})\Bigl(\frac{\tilde{\nu}}{\nu}\Bigr) ~\times 
        \frac{1}{4\Delta\nu} f_{21}(\nu)\times\cos\left(4\pi\nu t_\mathrm{cz}+2\phi\right),
\end{aligned}
\end{equation}
\noindent where $P(\nu)$ is a second order polynomial that removes the smooth contribution of the convective core from the signal \citep{cunha11}. The fitted quantities are the three second order polynomial parameters and $a_1$, $a_2$, $t_\mathrm{cz}$, and $\phi$. Here for convenience, we use the acoustic radius $t_\mathrm{cz}$, that is the counter part of the acoustic depth and is defined by changing the bounds of the integral in the left hand side of Eq.~\ref{eq:tau} to [0; $r_\mathrm{cz}$].

The fits of the signal in the frequencies and the second differences are performed similarly to Paper I with the code SIGS  (Seismic Inferences for Glitches in Stars) presented in \cite{pereira17}. Priors are applied for both $\tau_\mathrm{cz}$ and $\tau_\mathrm{HeII}$ such as $\tau_\mathrm{cz}>0.25\times\mathcal{T}>\tau_\mathrm{HeII}$. We checked that changing the transition from $0.25\times\mathcal{T}$ to lower values does not affect the results and conclusions of this work. As shown in paper I, from the theoretical point of view the ratio $a_1/a_2$ is really small. However, the data cannot lift the degeneracy between $a_1$ and $a_2$ when both terms are considered in the fitting, which led us to consider only the $a_2$ component for the frequency and the second differences. We do the same here and take into account only a single amplitude for the BSCZ signature in both indicators. We keep both terms for the ratios because they are the focus of this work. The fitted functions are
\begin{equation}\label{eq:fit_freq_std}
  \begin{aligned}
      \delta\nu =& A_\mathrm{cz}(\tau_\mathrm{cz}) \left(\frac{\tilde{\nu}}{\nu}\right) ~
      \cos(\Phi_{cz}) +  \delta\nu_\mathrm{HeII} ,
  \end{aligned}    
\end{equation}
where
\begin{equation}
 \delta\nu_\mathrm{HeII}  = A_\mathrm{HeII} ~\left(\frac{\tilde{\nu}}{\nu}\right) ~ \sin^2(2\pi\beta_\mathrm{HeII}\nu) ~\cos(4\pi\nu \tau_\mathrm{HeII}+2\phi_\mathrm{HeII})
\end{equation}
 
\begin{equation}\label{eq:fit_sd_std}
  \begin{aligned}
      \Delta^2\nu =& A_\mathrm{cz}^\ast(\tau_\mathrm{cz})\left(\frac{\tilde{\nu}}{\nu}\right) \sin \Phi_{cz}  +  \Delta^2\nu_\mathrm{HeII},
  \end{aligned}    
\end{equation}
\noindent where 
\begin{equation}
\Delta^2\nu_\mathrm{HeII}  = A_\mathrm{HeII}^\ast\left(\frac{\nu}{\tilde{\nu}}\right)\exp\left[-\beta_\mathrm{HeII}\left(\frac{\tilde{\nu}}{\nu}\right)^2\right]\sin(4\pi\nu \tau_\mathrm{HeII}+2\phi_\mathrm{HeII})
\end{equation}
and 
\noindent where $A_\mathrm{HeII}$ and $A_\mathrm{HeII}^\ast$ are the amplitude of the helium second ionisation region signature, $\beta_\mathrm{HeII}$ a constant, $\tau_\mathrm{HeII}$ the acoustic depth of the HeII region, and $\phi_\mathrm{HeII}$ a constant phase. The amplitudes $A_\mathrm{cz}$ and $A_\mathrm{cz}^\ast$ of the signature of the convective envelope can be expressed as function of $a_1$ and/or $a_2$ as presented in Eqs.~\ref{eq:mario}, \ref{eq:signal_r010}, and \ref{eq:signal_sd}. The fitted quantities are $A_\mathrm{cz}$, $A_\mathrm{HeII}$, $\beta_\mathrm{HeII}$, $\tau_\mathrm{cz}$, $\tau_\mathrm{HeII}$, $\Phi_\mathrm{cz}$, and $\phi_\mathrm{HeII}$ for Eq.~\ref{eq:fit_freq_std}. For  Eq.~\ref{eq:fit_sd_std} it is $A_\mathrm{cz}^\ast$, $A_\mathrm{HeII}^\ast$, $\beta_\mathrm{HeII}$, $\tau_\mathrm{cz}$, $\tau_\mathrm{HeII}$, $\Phi_\mathrm{cz}$, and $\phi_\mathrm{HeII}$.
 
The results we obtain correspond to what one gets with the method commonly used to fit the glitch signatures on the BSCZ. The distributions of the acoustic depth of the glitch are presented in Fig.~\ref{fig:tests_KIC667} (top panel) with the light grey histograms. For the frequencies and second differences, the distributions are not trivial to interpret, contrary to the one obtained from the ratios, which shows a single dominant peak.

\subsection{The sets of synthetic data} 
To better understand the distributions obtained from the observations, we generated sets of synthetic data in which we control the glitch signature. For that, we added the glitch signature to the zeroth order asymptotic expression for the frequencies, which is given by
\begin{equation}\label{nuas}
    \nu_{nl}=(n+\frac{l}{2}+\frac{1}{4}+\alpha)~\Delta\nu(n),
\end{equation}
\noindent where $n$ is the radial order, $l$ the angular degree, $\alpha$ a phase and $\Delta\nu(n)$ is the large frequency separation. For convenience, $\Delta\nu(n)$ is approximated with the mean large frequency separation $\Delta\nu$.
We modelled the frequency variation induced by the glitch of the BSCZ with a single simplified cosine term: 
\begin{equation}\label{eq:L}
    \begin{aligned}
    \delta\nu_\mathrm{cz}(\nu)\approx~~&A_2\left(\frac{\tilde{\nu}}{\nu}\right)\cos(4\pi\nu\tau_\mathrm{cz}+2\phi_1)\\
    \end{aligned},    
\end{equation}
\noindent with $A_2$ as a constant amplitude and $\phi_1$ as a constant phase. This contribution of the glitch is added to the asymptotic frequencies given by Eq.~\ref{nuas}.

In order to build a set of synthetic frequencies, we need to specify the values for the input parameters for the synthetic data. We arbitrary choose $A_2=0.15$~$\mu$Hz, $\phi_1=\pi$, and $\alpha=1.45$. Changing these values does not affect the results of the following analysis. $\tilde{\nu}=\nu_\mathrm{max}=941.8~\mu$Hz and $\Delta\nu=50.601~\mu$Hz are chosen to be the observed values for KIC6679371. We also need to give an input value to $\tau_\mathrm{cz}$. Among the several peaks seen in the observed distributions of $\tau_\mathrm{cz}$, we selected three that seem to be the most relevant: the two peaks around $\tau_\mathrm{cz}=2600$~s ($\tau_\mathrm{cz}/\mathcal{T}=0.26$) and $5000$~s ($\tau_\mathrm{cz}/\mathcal{T}=0.50$) in the distribution obtained from the second differences and the peak around $\tau_\mathrm{cz}=8300$~s ($\tau_\mathrm{cz}/\mathcal{T}=0.84$) in the distribution obtained from the ratios that seem to correspond to the one around $7900$~s ($\tau_\mathrm{cz}/\mathcal{T}=0.80$) in the distribution obtained from the frequencies. Selecting the third peak would not change the conclusions of the analysis. We discarded any peak below $2000$~s ($\tau_\mathrm{cz}/\mathcal{T}=0.20$), as it is unlikely that they are induced by the glitch at the position of the convective envelope.

For each value of a selected $\tau_\mathrm{cz}$, we generated a synthetic frequency set using Eq.~\ref{nuas}  for the zeroth-order asymptotic frequencies, and  Eq.~\ref{eq:L} to include a glitch signature at the desired value of $\tau_\mathrm{cz}$, for $l=0$, $1$, and $2$ modes. The Fourier transforms of the three synthetic sets for the three seismic indicators are computed using a Lomb-Scargle \citep{lomb76,scargle82} algorithm\footnote{Python package: scipy.signal.lombscargle} to identify the periodicity of the glitch signatures. The results are compared to the distributions obtained from the observation. We do not compare with the Fourier transform of the observed data because, contrary to the synthetic data, the observations include a smooth component and the contribution of the helium second ionisation region, which would pollute the comparison.

If we assume the peak at $\tau_\mathrm{cz}/\mathcal{T}\sim 0.50$ (i.e. around $5000$~s) in the distribution obtained from the second differences to be the real position of the BSCZ, the same peak is not found in the two other indicators (top panels of Fig.~\ref{fig:tests_KIC667}). The peak at 
$\tau_\mathrm{cz}/\mathcal{T}\sim 0.26$ (i.e. around $2600$~s) from the same distribution is somehow found in the distribution obtained from the frequencies, but not in the ratios (middle panels of Fig.~\ref{fig:tests_KIC667}). Finally, if we consider the peak at  $\tau_\mathrm{cz}/\mathcal{T}\sim 0.84$ (i.e. around $8300$~s) found in the distribution obtained from ratios (with a possible counterpart in the one obtained from the frequencies around $7900$~s) to be the real position of the BSCZ, then the signature cannot be found in the second differences (bottom panels of Fig.~\ref{fig:tests_KIC667}). 
 
\subsection{Sensitivity of the indicators}\label{sensi}

Seismic indicators are not sensitive to the same region of the star by construction, which could be the reason why we cannot find a common solution between the three indicators. In order to assess the sensitivity of each indicator to the region of the BSCZ, we consider the ratio amplitude-to-uncertainty $A/\sigma$. 

The amplitude $A$, is the amplitude of the signature in a given indicator estimated at $\nu =\tilde \nu=\nu_{max}$. From the theoretical expressions Eq.~\ref{eq:signal_r010} -using only the $a_2$ term- and Eq.~\ref{eq:signal_sd}), we obtained the ratio between A and $A_\nu$ for both the ratios and second differences as follows: 
\begin{eqnarray}
\frac{A_{r010}}{A_\nu } &=&   
        \frac{f_{21}(\tilde \nu)}{4\Delta\nu} ~~~~~;~~~~~~
\frac{A _{\delta_2\nu}}{A_\nu} =    \mathrm{ff}_{21}( \tilde \nu),
\end{eqnarray}
where $A_\nu$, $A_{r010}$, and $A _{\delta_2\nu}$ are the amplitude of the signature for the frequencies, second differences, and ratios, respectively. For high order p modes ($\nu >> \Delta \nu$) (from paper I), we have
\begin{eqnarray}
f_{21} &\approx&  \cos (2 \Phi_d)-4\cos(\Phi_d) +3\\
\mathrm{ff}_{21}  &\approx&  -\sin^2 \Phi_d \\
\Phi_d &=& 2\pi~ \Delta \nu ~\tau_d
.\end{eqnarray}

\noindent The $\sigma$ was determined by the mean of the observed uncertainties of the seismic indicator (frequencies, second differences, or ratios) between $0.8\times \nu_\mathrm{max}<\nu<1.2\times \nu_\mathrm{max}$ for the reference star. We then estimated the ratios 
$${\cal R}_1 = \frac{A_{r010}/\sigma_{r010}}{A_\nu/\sigma_\nu} ~~~~~;~~~{\cal R}_2 = \frac{A_{\delta_2\nu}/\sigma_{\delta_2\nu}}{A_\nu/\sigma_\nu}$$ for several values of the normalised acoustic depth of the BSCZ $\tau_{cz}/\mathcal{T}$.  
 
Figure~\ref{fig:sensi} represents the ratios ${\cal R}_1$ and ${\cal R}_2$ as a function of $\tau_{cz}/\mathcal{T}$. It shows that the second differences are more sensitive to the centre of the acoustic cavity while the $r_{010}$ ratios are more sensitive to the core of the star. This is the reason why the glitch signature can be more easily detected in one indicator than the other depending on the (acoustic) position of the BSCZ in the star. 

The fact that the frequencies and ratios give different results than the second differences could have different explanations. Firstly, the second differences and ratios could highlight glitches of different origin, one coming from the convective envelope (second differences), and the other one (ratios) coming from a sharp variation of the sound speed deeper inside the star (e.g. strong chemical gradient). Except for convective boundaries, there is currently no theoretical explanation for a sharp variation of the sound speed located below the second ionisation region of helium from the thermal, thermodynamical, or chemical stratification. However, in this case the peak at $\tau_\mathrm{cz}/\mathcal{T}\sim 0.84$ (i.e. around $8300$~s) in the distribution obtained from the ratios cannot be related to any of the convective boundaries. From stellar structure models we expect the bottom of the convective envelope to be around $\tau_\mathrm{cz}/\mathcal{T}\sim 0.4$ and the top of the convective core to be of the order of $\tau_\mathrm{cz}/\mathcal{T}\sim 0.02$.

Secondly, while it is not possible to confuse acoustic radius and depth with the ratios (aliasing described by \citealt{mazumdar01}) because of the asymmetry of the sensitivity of the indicator, it is possible that such confusion arises for the other two indicators. The peak around $\tau_\mathrm{cz}/\mathcal{T}\sim 0.25$ seen in the distribution obtained from the second differences could then be confused with a peak around $\tau_\mathrm{cz}/\mathcal{T}\sim 0.75$, which could solve the problem. However, the  position of the glitch would still remain a problem (too deep compared to theoretical expectations).

Thirdly, the uncertainties on the individual frequencies prevent the proper measurement of the glitch signature if the amplitude of the signal is not large enough. This could explain why the distributions obtained from the frequencies and the second differences are more difficult to interpret. But it does not explain why the signature in the ratios indicates such a deep glitch, assuming it is linked to the convective envelope.

Given the above, in the present study, we rather explore an alternative explanation where both seismic indicators show the same glitch -the glitch from the bottom of the convective envelope- but we miss terms in the usual theoretical expression for the glitch signature. In that case, we could therefore reasonably assume that the signature is actually non-sinusoidal (which seems to be the case for KIC6679371, see Fig.~1 of Paper I) because of the contribution of additional non-linear terms or higher order terms in the perturbative (asymptotic) approach. We then expect that the first additional contribution comes from a term with twice the acoustic depth (see e.g. \citealt{provost93}), corresponding to $\tau_\mathrm{cz}/\mathcal{T}=0.80\pm 0.10$ (i.e. $8000\pm1000$~s), and the real signature of the BSCZ would correspond to $\tau_\mathrm{cz}/\mathcal{T}\sim 0.40\pm 0.10$ (i.e. around $4000\pm1000$~s). We note that, this explanation is not data-driven since the current fitting expression already reproduces the data (see Appendix~\ref{Apdx:fits}) because of the large uncertainties on the frequencies for F-type stars. It should rather be seen as an exploratory study based on a theoretical attempt to explain the discrepancy between the data and the stellar structure models.

\subsection{A more complex signature of the glitch}

\begin{figure}
    \centering
    \includegraphics[scale=0.6]{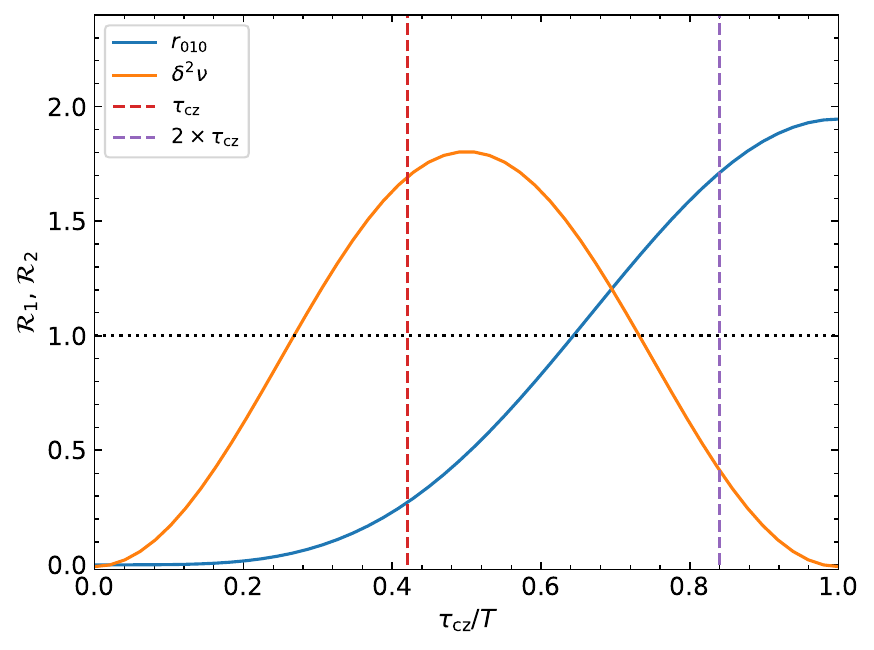}
   \caption{Signal-to-uncertainty ratio of the BSCZ signature (see the text for the definition) in the $r_{010}$ ratios (blue curves) and the second differences (orange curves) normalised by the signal-to-uncertainty ratio observed in the frequency, according to the scaled acoustic depth of the BSCZ for KIC6679371. We note that from one star to another, this figure is very similar. The horizontal dotted line indicates where the signal-to-noise ratio of the signature of the BSCZ is equal to the signal-to-noise ratio of the frequencies. Both vertical dashed lines indicate  the position of the $\tau_\mathrm{cz}$ and $2\times\tau_\mathrm{cz}$ inputted in the synthetic data.}
   \label{fig:sensi}
\end{figure}

To test the possibility of a more complex signature of the glitch, we generated an new set of synthetic frequencies using Eq.~\ref{eq:NL}. Here we add an additional component to the above standard term of the glitch signature (Eq.~\ref{eq:L}), according to
\begin{equation}\label{eq:NL}
    \begin{aligned}
    \delta\nu_\mathrm{cz}(\nu)\approx~~&A_2 ~\left(\frac{\tilde{\nu}}{\nu}\right)~\cos(4\pi\nu\tau_\mathrm{cz}+2\phi_1)\\
    + & ~~k \times A_2 ~\left(\frac{\tilde{\nu}}{\nu}\right)~ \cos(8\pi\nu\tau_\mathrm{cz}+2\phi_2)
    \end{aligned},    
\end{equation}
\noindent with $k$ as a multiplicative factor and $\phi_2$ as a constant phase. The second term is equivalent to a signal with twice the acoustic depth of the first order term. We note that $\phi_2$ was not set to $2\times\phi_1$ in order to have more freedom in the synthetic data. The parameters of the synthetic glitches are chosen to best reproduce the major peak of the distribution of $\tau_\mathrm{cz}$ and $t_\mathrm{cz}$ obtained from the three indicators of the observed stars. 

We tested again all the peaks in the distributions and selected only the values of $\tau_\mathrm{cz}$ giving the best agreement with the observed distributions, meaning ($\tau_\mathrm{cz}/\mathcal{T}=0.42$ and $\tau_\mathrm{cz}/\mathcal{T}=0.84$; $\tau_\mathrm{cz}=4150$~s and $8300$~s, respectively). However, the analysis is first performed with the first value. Hence, we use $A_2=0.15~\mu$Hz, $\tau_\mathrm{cz}/\mathcal{T}=0.42$  (half of the peak in the ratios) (i.e. $\tau_\mathrm{cz}=4150$~s), $k=1$ and  $\phi_1=\phi_2=\pi$. The Fourier transforms are shown in the top panels of Fig.~\ref{fig:tests_KIC667_ns}. The reader must not pay attention to the amplitudes of the Fourier peaks. Indeed, the Fourier transforms presented in the Fig.~\ref{fig:tests_KIC667_ns} are sometimes re-scaled for clarity, only the positions of the Fourier peaks matter. It can be seen that the period of the signature in the second differences and the ratios are well explained by construction of the synthetic data. Only one of the two periods is seen in the second differences because the sensitivity of this indicator is maximum around $\mathcal{T}/2$ and minimum at the edge of the acoustic cavity (see Fig.~\ref{fig:sensi}). Similarly, only one period is seen in the ratios because this seismic indicator is mainly sensitive to signals with a large acoustic depth (see Fig.~\ref{fig:sensi}). However, both periods are seen in the frequencies as expected from the construction of the synthetic data. We performed the same tests with $k=5$, meaning that the additional term has a much larger amplitude than the standard term. As seen in the middle panel of Fig.~\ref{fig:tests_KIC667_ns}, the distributions obtained for the three indicators are consistent with the periodicities found in the synthetic data (with a slight shift for the frequencies). However, the peak around $\tau_\mathrm{cz}/\mathcal{T}=0.84$ in the second differences is not consistent with the observed distribution.

The bottom panel shows a case with $k=3$ and a random noise (Gaussian with a standard deviation of $0.25~\mu$Hz representative of KIC6679371) is added to the frequencies. In this case we can clearly see that the distributions obtained from the frequencies and ratios are almost unaffected, whereas the second differences are affected. In that case, the distribution obtained with the second differences is in better agreement with the one obtained from the observations (no clear peaks). This is an indication that in such a case, the ratios are probably more reliable than the second differences to measure the position of the BSCZ.   

With this analysis, we show that if the signature seen in the three seismic indicators are all related to the BSCZ, we can only get a good agreement with an additional term in the theoretical glitch signature expression. In this case, only one of the two solutions is acceptable (i.e. the one giving the shallowest convective envelope $\tau_\mathrm{cz}/{\cal \tau}=0.42$; 4150~s) because the other solution ($\tau_\mathrm{cz}/{\cal \tau}=0.84$; 8300 ~s) is way too deep compared with what we could expect from stellar models but not deep enough to correspond to a glitch localised in the very central layers. This would correspond to a convective envelope more than twice the size predicted by stellar model, with an increase of more than $4~H_p$. If this increase is caused by penetrative convection, 3D simulations only predict about $\approx0.3~H_p$ \citep{breton22} for F-type stars.

As a conclusion of this analysis, it appears that some of the peaks seen in the distributions obtained from the three seismic indicators cannot be identified when using the standard expression for the glitch signature of the BSCZ. The standard expression is therefore not able to catch the complex (non-sinusoidal) shape of the signature found for the F-type stars. In contrast,  adding an extra term with twice the acoustic depth of the standard expression can reconcile the results obtained for the three seismic indicators. We shall explore further this possibility in the remaining of the paper.

\section{Procedure to test the possibility of complex signatures of the glitch}\label{sect:synth}

\begin{table}
\centering
\caption{Properties of the glitches included in the synthetic data.} 
\label{tab:1}
\resizebox{\columnwidth}{!}{%
\begin{tabular}{lccccccc}
\hline\hline
KIC & Group & $\tau_\mathrm{cz}$ [s] & $\tau_\mathrm{cz}/\mathcal{T}$ & $k$ & $A_2$ [$\mu$Hz] & $\phi_1$ & $\phi_2$  \\
\noalign{\smallskip}\hline 
1435467  & 1 & $4250$ & $0.60$ & -     & $0.15$ & $\pi/4$ & -         \\
1435467  & 2 & $4250$ & $0.60$ & $2.0$ & $0.15$ & $2.2$   & $1.1$       \\
\hline
2837475  & 1 & $2700$ & $0.41$ & -     & $0.35$ & $\pi$   & -         \\
2837475  & 2 & $2700$ & $0.41$ & $4.0$ & $0.15$ & $2.2$   & $1.1$       \\
\hline
6679371  & 1 & $2600$ & $0.26$ & -     & $0.15$ & $\pi$   & -         \\
6679371  & 2 & $4250$ & $0.43$ & $3.0$ & $0.15$ & $2.2$   & $1.1$    \\
\hline
9206432  & 1 & $2900$ & $0.49$ & -     & $0.35$ & $\pi$   & -         \\
9206432  & 2 & $1430$ & $0.24$ & $4.0$ & $0.15$ & $2.2$   & $1.1$    \\
\hline
10162436 & 1 & $5400$ & $0.60$ & -     & $0.15$ & $\pi$   & -         \\
10162436 & 2 & $6250$ & $0.70$ & $1.0$ & $0.15$ & $2.2$   & $1.1$ \\
\hline
11081729 & 1 & $2200$ & $0.40$ & -     & $0.35$ & $\pi$   & -         \\
11081729 & 2 & $1700$ & $0.31$ & $3.0$ & $0.25$ & $2.2$   & $1.1$    \\
\hline
11253226 & 1 & $3800$ & $0.58$ & -     & $0.35$ & $\pi$   & -         \\
11253226 & 2 & $1900$ & $0.29$ & $4.0$ & $0.25$ & $2.2$   & $1.1$    \\
\hline
12317678 & 1 & $2700$ & $0.34$ & -     & $0.35$  & $\pi$   & -         \\
12317678 & 2 & $2700$ & $0.34$ & $1.0$ & $0.25$ & $2.2$   & $1.1$    \\
\noalign{\smallskip}\hline\noalign{\smallskip}
\end{tabular}
}%
\end{table}

Our goal is to test the possibility that the signature of the glitch for F-stars is actually non-sinusoidal. This implies that additional terms must be added to the theoretical expression for the glitch signature to account for the observed non-sinusoidal shape of the glitch signature in F-type stars. To test that hypothesis, we start by carrying out fits of the three indicators for the observed stars assuming either a standard expression for the glitch signature (Eq.~\ref{eq:L}) or a non-sinusoidal expression (Eq.~\ref{eq:NL}). In a second step and in order to help the interpretation of those distributions, we generated sets of synthetic frequencies that include different expressions for the glitch signatures. 

\subsection{Target selection}\label{sec:samples}

\begin{figure}
    \centering
    \includegraphics[scale=0.6]{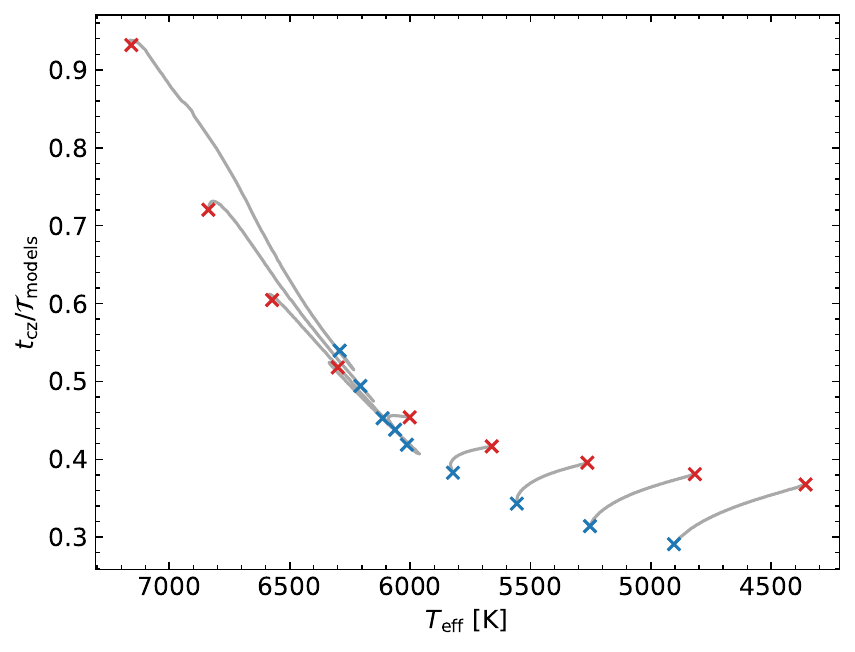}
    \includegraphics[scale=0.6]{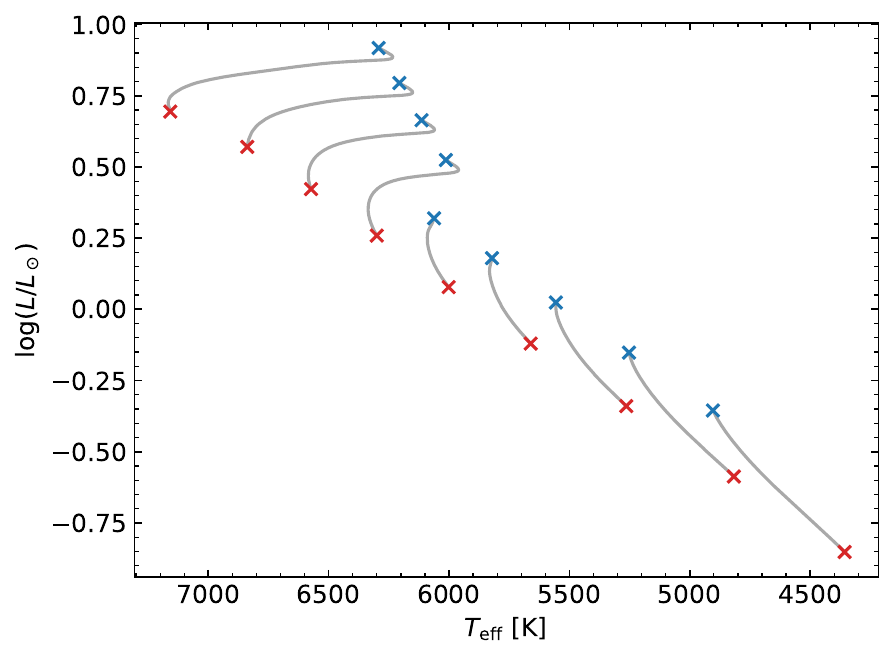}
   \caption{Predictions from stellar models. \textit{Top panel:} Variation of the ratio between the acoustic radius of the BSCZ and the total acoustic radius for stellar models between the zero age main sequence (red crosses) and the terminal age main sequence (blue crosses). The masses vary between $0.7$ and $1.5$~M$_\odot$. \textit{Bottom panel:} HR diagram for the same models.}
   \label{fig:predi}
\end{figure}

The position of the BSCZ is strongly correlated with the effective temperature of the star. We show in Fig.~\ref{fig:predi} that the variation of the acoustic radius of the BSCZ according to the effective temperature is different for a G- and a F-type star. For all the cases, the BSCZ deepens with the evolution but the effective temperature increases for G-type stars while it decreases for  F-type stars. The transition between the two types of behaviours occurs when the CNO cycle starts to dominate the energy production (i.e. when the star develops a convective core). This transition occurs around $T_\mathrm{eff}\approx6100$~K. We selected the targets following this distinction. The analysis was performed for eight stars: 
\begin{itemize}
    \item Two G-type stars with $T_\mathrm{eff}<5800$~K, namely the Sun and Doris (KIC8006161);
    \item The six hottest \textit{Kepler} stars with the best signal-to-noise ratios, KIC2837475, KIC6679371, KIC9206432, KIC11081729, KIC11253226, and KIC12317678 ($T_\mathrm{eff}>6400$~K), hereafter called sample A; and
    \item Two stars in the intermediate regime $5900<T_\mathrm{eff}<6400$~K, KIC1435467 and KIC10162436, hereafter called sample B.
\end{itemize}
Three of the stars (i.e. KIC1435467, KIC6679371, and KIC10162436) were already identified in Paper I as interesting targets. In the following, we describe how the synthetic data are generated for the F-type stars. No synthetic data were necessary for the G-type stars because their analysis is straightforward.

\subsection{Sets of synthetic frequencies with a standard glitch signature (Group 1)}
\begin{figure*}
    \centering
    \includegraphics[scale=0.65]{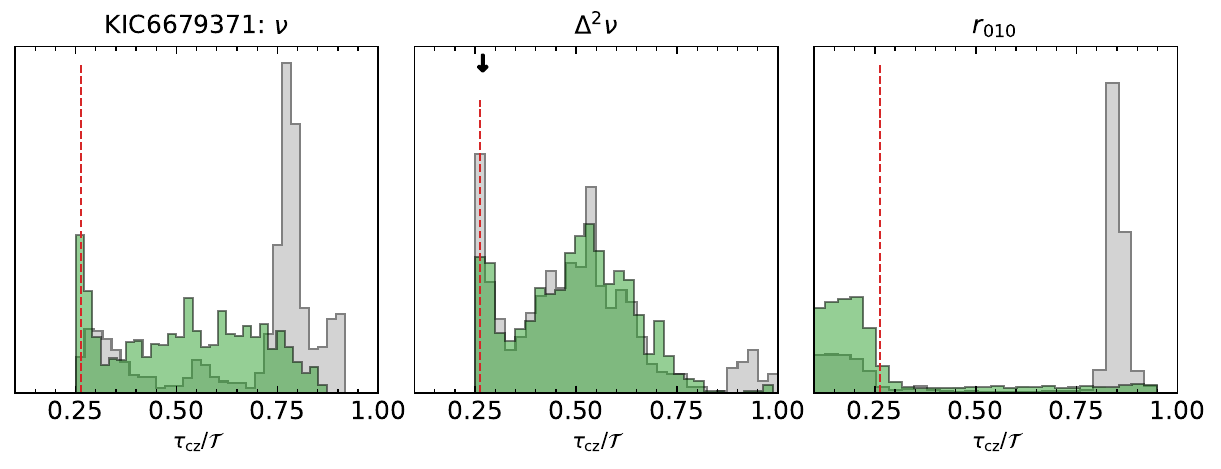}
    \includegraphics[scale=0.65]{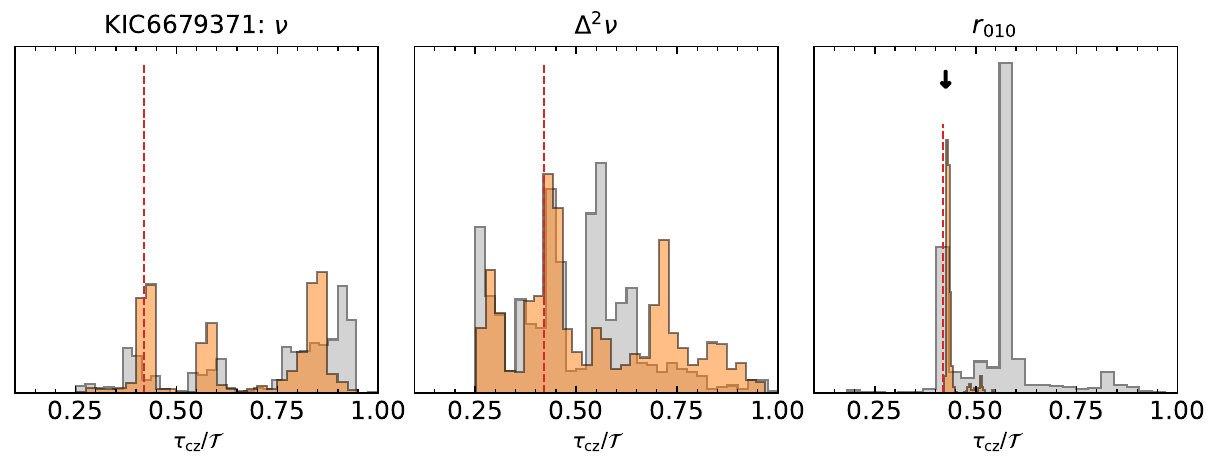}
   \caption{Distributions of $\tau_\mathrm{cz}/\mathcal{T}$ for the glitch signature in the frequencies (left panels), second differences (middle panels), and $r_{010}$ ratios (right panels) for the \textit{Kepler} star KIC6679371. The results for the $r_{010}$ ratios are converted from the measured $t_\mathrm{cz}$ with $\tau_\mathrm{cz}$= $T_0-t_\mathrm{cz}$, $T_0=1/(2\Delta\nu)$ being the total acoustic radius of the star. The background grey histograms are obtained from the observations and the transparent coloured histograms (green, blue, and orange, from top to bottom) are obtained from the synthetic data. The top panels are obtained by a fit of the observed and synthetic data of Group 1 with the standard expressions (Eq.~\ref{eq:fit_r010_std}, \ref{eq:fit_freq_std}, \ref{eq:fit_sd_std}). The bottom panels are obtained by a fit of the observed and synthetic data of Group 2 with the expressions that included the additional term (Eq.~\ref{eq:ns_r010} and \ref{eq:ns_freq_sd}). The red dashed vertical lines indicate the position of the BSCZ as inputted in the synthetic data. The
black down arrows show the dominant peaks of the grey distributions shown in top panel.}
   \label{fig:KIC667}
\end{figure*}

Similarly to the first set of synthetic data presented in Sect.~\ref{sec:issue}, we model the frequency variation induced by the glitch of the BSCZ with a single cosine term (Eq.~\ref{eq:L}). This contribution of the glitch is added to the asymptotic frequency (Eq.~\ref{nuas}) using the values of $\Delta\nu$ of the reference stars. The parameters of the synthetic standard glitches are chosen to best reproduce the peak of highest amplitude in the distribution of $\tau_\mathrm{cz}$ obtained from a standard fit of the second differences of the observed stars (upper middle panel of Figs.~\ref{fig:KIC667}, \ref{fig:KIC143}, and \ref{fig:KIC1016}). We used the second differences as reference because this is the most commonly used seismic indicator for the measurement of the position of the BSCZ. Similarly to Sect.~\ref{sec:issue}, whichever peak is selected, the conclusions remain the same. We chose $\tilde{\nu}=\nu_\mathrm{max}$ and the other parameters are presented in Table~\ref{tab:1}. To each synthetic frequency, we associated an error bar given by that of the frequency of the corresponding mode ($l$ and $n$) of the reference star in order to be as close as possible to a realistic signal.

\subsection{Sets of synthetic frequencies with a non-sinusoidal expression of the glitch signature (Group 2)}
Similarly to the second set of synthetic data presented in Sect.~\ref{sec:issue}, we add an additional component to the above standard term of the glitch signature (Eq.~\ref{eq:L}), according to Eq~\ref{eq:NL}. This additional term is equivalent to a signal with twice the acoustic depth of the first order term. The parameters of the synthetic glitches are chosen to best reproduce the main peak of the distribution of $\tau_\mathrm{cz}$ obtained from the three indicators of the observed stars (prioritising the ratios if a choice needs to be made). They are presented in Table~\ref{tab:1}.\\

\begin{table*}
\centering
\caption{Properties of the selected targets.} 
\label{tab:2}
\begin{tabular}{lccccccccc}
\hline\hline
KIC & $T_\mathrm{eff}$ [K] & $\nu_\mathrm{max}$ [$\mu$Hz] & $\Delta\nu$ [$\mu$Hz] & $\tau_\mathrm{cz,std}$ [s] & $\tau_\mathrm{cz,ns}$ [s] & $t_\mathrm{cz,std}/\mathcal{T}$ & $t_\mathrm{cz,ns}/\mathcal{T}$ & $\tau_\mathrm{cz,std}/\mathcal{T}$ & $\tau_\mathrm{cz,ns}/\mathcal{T}$ \\
\noalign{\smallskip}\hline
2837475  & $6614\pm77$ & $1557.6^{+8.2}_{-9.2}$  & $75.729^{+0.034}_{-0.033}$ & $1999^{+236}_{-206}$ & $2676^{+269}_{-290}$\tablefoottext{a} &$0.697^{+0.031}_{-0.036}$ & $0.595^{+0.044}_{-0.041}$ &$0.303^{+0.036}_{-0.031}$ & $0.405^{+0.041}_{-0.044}$\\
6679371  & $6479\pm77$ & $941.8^{+5.1}_{-5.0}$  & $50.601^{+0.029}_{-0.029}$ & $8397^{+207}_{-179}$ & $4168^{+165}_{-92}$ &$0.150^{+0.018}_{-0.021}$ & $0.578^{+0.009}_{-0.016}$ &$0.850^{+0.021}_{-0.018}$ & $0.422^{+0.016}_{-0.009}$   \\
9206432  & $6538\pm77$ & $1866.4^{+10.3}_{-14.9}$  & $84.926^{+0.046}_{-0.051}$ & $2869^{+85}_{-70}$ & $1430^{+40}_{-31}$ &$0.513^{+0.012}_{-0.015}$ & $0.757^{+0.005}_{-0.007}$ &$0.487^{+0.015}_{-0.011}$ & $0.243^{+0.007}_{-0.005}$ \\
11081729  & $6548\pm82$ & $1968.3^{+11.0}_{-12.6}$ & $90.116^{+0.048}_{-0.047}$ & $2180^{+162}_{-142}$ & $1727^{+43}_{-130}$ &$0.607^{+0.026}_{-0.029}$ & $0.689^{+0.023}_{-0.008}$ &$0.393^{+0.029}_{-0.026}$ & $0.311^{+0.008}_{-0.026}$ \\
11253226  & $6642\pm77$ & $1590.6^{+10.6}_{-6.8}$ & $76.858^{+0.026}_{-0.030}$ & $3820^{+72}_{-79}$ & $1881^{+40}_{-50}$ &$0.413^{+0.012}_{-0.011}$ & $0.711^{+0.008}_{-0.006}$ &$0.587^{+0.011}_{-0.012}$ & $0.289^{+0.006}_{-0.008}$ \\
12317678  & $6580\pm77$ & $1212.4^{+5.5}_{-4.9}$ & $63.464^{+0.025}_{-0.024}$ & $2376^{+255}_{-255}$ & $2606^{+95}_{-85}$\tablefoottext{a} & $0.698^{+0.032}_{-0.032}$ & $0.669^{+0.011}_{-0.012}$  & $0.302^{+0.032}_{-0.032}$ & $0.331^{+0.012}_{-0.011}$\\
\noalign{\smallskip}\hline
1435467  & $6326\pm77$ & $1406.7^{+6.3}_{-8.4}$ & $70.369^{+0.034}_{-0.033}$  & $5317^{+149}_{-238}$ & $4370^{+76}_{-80}$ & $0.252^{+0.033}_{-0.021}$ & $0.385^{+0.015}_{-0.016}$  & $0.748^{+0.021}_{-0.033}$ & $0.615^{+0.016}_{-0.015}$ \\
10162436 & $6146\pm77$ & $1052.0^{+4.0}_{-4.2}$ & $55.725^{+0.035}_{-0.039}$ & $5234^{+137}_{-281}$ & $6261^{+80}_{-90}$ & $0.418^{+0.031}_{-0.015}$ & $0.302^{+0.010}_{-0.009}$ & $0.582^{+0.015}_{-0.031}$ & $0.698^{+0.009}_{-0.010}$ \\
- & - & - & - & $6395^{+223}_{-171}$ & $6261^{+80}_{-90}$ & $0.287^{+0.019}_{-0.025}$ & $0.302^{+0.010}_{-0.009}$ & $0.713^{+0.025}_{-0.019}$ & $0.698^{+0.009}_{-0.010}$ \\
\noalign{\smallskip}\hline
8006161 & $5488\pm77$ & $3574.7^{+11.4}_{-10.5}$ & $149.427^{+0.015}_{-0.014}$ & $2245^{+35}_{-56}$ & $2230^{+84}_{-61}$ & $0.329^{+0.017}_{-0.011}$ & $0.334^{+0.018}_{-0.025}$ & $0.671^{+0.011}_{-0.017}$ & $0.666^{+0.025}_{-0.018}$ \\
Sun & $5777\pm3$ & $3090$ & $135.1$ & $2273^{+12}_{-12}$ & $2271^{+9}_{-8}$ & $0.386^{+0.003}_{-0.003}$ & $0.386^{+0.002}_{-0.002}$ & $0.614^{+0.003}_{-0.003}$ & $0.614^{+0.002}_{-0.002}$ \\
\noalign{\smallskip}\hline\noalign{\smallskip}
\end{tabular}
\tablefoot{The $T_\mathrm{eff}$, $\nu_\mathrm{max}$, $\Delta\nu$ values are taken from \cite{lund17}. The acoustic depths and radii are obtained by the fitting of the ratios $r_{010}$. \textit{std} stands for the standard fitting expression and \textit{ns} stands for the fitting with the non-sinusoidal expression. \tablefoottext{a}{Measured from the second differences distribution.}}
\end{table*}

From these six sets of frequencies, we computed the second differences $\delta^2\nu$ and the ratios $r_{010}$ following the expressions in Eq.~\ref{eq:sec_dif} and \ref{eq:r010}, respectively. We then fit the synthetic data built using the parameters characteristics of the three stars using the same procedure as for the observations.

\subsection{Standard and non-sinusoidal glitch signature expressions}

Firstly, the fits of the observed and synthetic data (Group 1) are performed using the standard expressions for the glitch signatures with Eq.~\ref{eq:fit_freq_std}, \ref{eq:fit_sd_std}, and \ref{eq:fit_r010_std} (standard fit, see Sec~\ref{sec:issue}). The results we obtained correspond to what one gets with the method commonly used to fit the glitch signatures on the BSCZ.

Secondly, the fits of the observed and synthetic data (Group 2) are performed with the addition of a term making the signature non-sinusoidal (hereafter called non-sinusoidal fit) of the form
\begin{equation}\label{eq:ns_r010}
    k \times A_2\left(\frac{\tilde{\nu}}{\nu}\right)\cos(4\pi\nu(\mathcal{T}-2\tau_\mathrm{cz})+2\phi_2)
\end{equation}

\noindent for the ratios (because what is measured by the $r_{010}$ ratios is the acoustic radius and not the acoustic depth, but the factor of two is on the acoustic depth) and
\begin{equation}\label{eq:ns_freq_sd}
    k \times A_2\left(\frac{\tilde{\nu}}{\nu}\right)\cos(8\pi\nu\tau_\mathrm{cz}+2\phi_2)
\end{equation}
\noindent for the frequencies and second differences.

In the following sections, we compare the observed and synthetic distributions in a single plot for each star and each indicator. This enables us to correctly interpret the results of the fit of the synthetic data and at the same time obtain a possible interpretation for the peaks appearing in the real data. We first applied this procedure with the fit of the standard glitch signature formulation, and in a second step, we applied it to the glitch signature expression with an additional term.

\subsection{Degeneracy of the solution}
Before analysing the results of the above procedure, we must give a word of caution. A fitting expression that includes an additional term with twice the acoustic depth of the first term introduces a degeneracy in the fits for the $r_{010}$ ratios between values of $k>1$ and values of $k$ close to zero (similar to the standard case). In one case, we get a value of $\tau$ and in the other case the solution is twice this value. It can lead to double peak solutions or to the wrong solution being dominant (according to the fact that we know what is the right solution in the synthetic data). Most of the time, the wrong solution is deeper inside the star than the correct one, and for $t_\mathrm{cz}/\mathcal{T}<0.5$, that is much deeper than what we expect from theoretical stellar structure models (see Sect.~\ref{sect:comp_obs}). To avoid confusion, we discuss hereafter only values of $\tau_\mathrm{cz}/\mathcal{T} <0.5$ ($t_\mathrm{cz}/\mathcal{T}>0.5$).

\section{Distributions of the acoustic depth of the surface convective zone for the hottest stars (sample A)}\label{sec:sampleA}

In this section we analyse the 6 hottest stars of the \textit{Kepler} Legacy sample (sample A). The $\tau_\mathrm{cz}$ distributions are shown in Fig.~\ref{fig:KIC667} for KIC6679371, and in Figs. \ref{fig:KIC920}, \ref{fig:KIC110}, \ref{fig:KIC112} for KIC9206432, KIC11081729, and KIC11253226, respectively. Each star is analysed separately in the following subsections.

\subsection{KIC6679371} 
\paragraph{Synthetic data}

\textit{Standard fits:} The synthetic distributions of $\tau_\mathrm{cz}$ obtained for Group 1 using the standard fit are presented on the top panels. The synthetic distribution obtained with the fit of the second differences shows a peak at the inputted position of the BSCZ, an additional broad peak appears at the position of the middle of the acoustic cavity (around 5000~s). For the frequencies, a dominant peak is found in the distribution at the expected position. No clear dominant peak is found for the ratios. With a value of $\tau_\mathrm{cz}/\mathcal{T} \sim 0.15$ (Table~\ref{tab:2}) and looking at Fig.~\ref{fig:sensi}, a dominant peak in the $r_{010}$ distribution should not appear, which is expected because the ratios are not sensitive to this region of the star. 

\textit{Non-sinusoidal fits:} The fit of the Group 2 synthetic data with the expression including the additional term are presented in the bottom panels. The distributions in the frequencies and in the second differences display complex patterns that make difficult to derive the position of the BSCZ with those two indicators. In contrast, the distribution for the $r_{010}$ ratios is single peaked at the position of the BSCZ inputted in the data.

\paragraph{Observed data}

\textit{Standard fits:} By construction of the synthetic data, the observed distribution in the second differences is very similar to the one obtained from the synthetic data. But for the frequencies and the ratios, the dominant peaks are found at different locations. It is particularly striking for the ratio with a difference of about 6000~s, that corresponds to 60\% of the acoustic radius of the star.

\textit{Non-sinusoidal fits:} For each indicator, the observed distribution is compared to the synthetic distribution (group 2), both fitted with a non-sinusoidal formulation. Similar to the synthetic data, the observed distributions for the second differences and the frequencies show complex patterns that are not suitable for deriving  the value of the position of the BSCZ. We note, however, that both the synthetic and the observed patterns in the frequency distributions show a similar pattern with three dominant peaks. Again here, the distribution in the ratios is better defined with two dominant peaks. One of these peaks coincides with the dominant peak of the synthetic data at the exact position of inputted position of the BSCZ. It is important to note that among the two major peaks of the distribution, one is located at $t_\mathrm{cz}/\mathcal{T}<0.5$ (which we therefore does not consider) and the second one, around $4150$~s, is in better agreement with the prediction of stellar structure models (see Sect.~\ref{sect:comp_obs}).

\paragraph{Conclusion}
Assuming that one of the dominant peaks provides the position of the BSCZ, a coherent result (the same position of the dominant peak for all three indicators) is obtained only when the glitch signature is considered as non sinusoidal. In this case, the ratios are more reliable to measure the position of the BSCZ provided one includes an additional term with twice the acoustic depth of the actual position.

\subsection{KIC9206432} 

We now discuss the results obtained for the reference star KIC9206432 shown in Fig.~\ref{fig:KIC920}.

\textit{Standard fits:} Unlike the case of KIC6679371, the Group 1 synthetic distributions obtained for the three indicators provide the same measurement of the position of the BSCZ at the expected position. By construction of the synthetic data, the observed distributions for the three seismic indicators are respectively similar to the ones obtained from the synthetic data, except for a second peak in the second differences.

\textit{Non-sinusoidal fits:} For each indicator respectively, the observed and Group 2 synthetic distributions, both fitted with a non-sinusoidal formulation, show a similar pattern. However, only the ratio-based distributions provide a single unambiguous peak that moreover corresponds to the correct position of the BSCZ inputted value in the synthetic data.

\paragraph{Conclusion} 
The synthetic and observed distributions show a similar pattern respectively for each indicator. When including an additional term with twice the acoustic depth of the actual position, the most reliable measurement of the position of the BSCZ is given by the ratios.

\subsection{KIC11081729} We now discuss the results obtained for the reference star KIC11081729 in Fig.~\ref{fig:KIC110}.

\textit{Standard fits:} The synthetic distributions for the frequencies and the second differences show a pattern with two regions of dominant peaks. In each case, one of the two regions is found to be centred around the inputted position of the BSCZ. The dominant peak in the ratio distribution is found at the inputted acoustic depth. By construction of the synthetic distributions, the observed distributions for the three seismic indicators are very similar to the synthetic ones, except for an extra peak around $4200~s$ in the three observed distributions.

\textit{Non-sinusoidal fits:} The observed and Group 2 synthetic distributions look very similar to the standard fit ones, except that the position of the BSCZ as given by the ratio single peak is shifted to a slightly lower value than expected.

\paragraph{Conclusion} 
In this case, the three indicators give the same results independently of the fitting expression. The non sinusoidal fits of the synthetic data reproduce slightly better the observed ones. We stress that the distributions obtained from the ratios allow for a better measurement of the position of the BSCZ than the other two indicators. The value of $\tau_\mathrm{cz}$ given by the non-sinusoidal fits are in slightly better agreement with the theoretical prediction (see Sect.~\ref{sect:comp_obs}).

\subsection{KIC11253226} 
We now discuss the results obtained for the reference star KIC11081729 in Fig.~\ref{fig:KIC112}.

\textit{Standard fits:} The synthetic distributions for the three indicators all provide the correct measurement of the position of the BSCZ without ambiguity. By construction of the synthetic data, the observed distributions for the three seismic indicators are very similar to the one obtained from the synthetic data, except for extra small amplitude peaks around the major peak in the frequencies and the second differences.

\textit{Non-sinusoidal fits:} The Group 2 synthetic and observed distributions obtained for the frequencies and the second differences are noisy and difficult to interpret, but a peak at the right position is present for both indicators. In contrast, the synthetic and observed ratio-based distributions show one single, unambiguous peak at the expected position.

\paragraph{Conclusion} 
Similar to the case of KIC6679371, the ratios give a clear measurement of the acoustic depth of the BSCZ whereas the other two indicators cannot really be interpreted. Also the ratios appear more reliable with a non-sinusoidal fit. Moreover, the measured position of the BSCZ is in better agreement with theoretical predictions (see Sect.~\ref{sect:comp_obs}).  

\subsection{KIC2837475 and KIC12317678}\label{sec:KIC283}

The last two stars of sample A are in the specific case where the position of the BSCZ cannot be measured from the ratios, but can be measured from the two other indicators. Indeed the position of the BSCZ is found to be of the order of $\tau_\mathrm{cz}/\mathcal{T} \sim 0.3-0.4$ (Table~\ref{tab:2}) where the ratios are much smaller than the second differences (see Fig.~\ref{fig:sensi}). 

In Figs.~\ref{fig:KIC283} and ~\ref{fig:KIC123}, the comparisons between observed and synthetic data show that there is a better agreement with a non-sinusoidal fit of the glitch signature for the distributions obtained from the frequencies and the second differences. For the ratios of the synthetic data of KIC2837475 (Group 2), we obtain a better agreement with the observed distribution when assuming a smaller $k$, despite providing a good measurement of the BSCZ (see the bottom panels of Fig.~\ref{fig:KIC283}). For KIC12317678, the difficulty to measure the position of the BSCZ seems to come from a too small amplitude of the glitch signature (see the bottom right panel of Fig.~\ref{fig:KIC123}). This comparison also allows the different peaks to be disentangled and for the determination that the position of the BSCZ corresponds to the peak around $2650$~s for both KIC2837475 and KIC12317678.

\subsection{Interpretation of the results}

The above analyses indicate that standard glitch fitting methods may not be adapted for the hotter F-type stars and could lead to wrong estimates of the position of the BSCZ. The results show that the signature of the ratios should be considered at least as reliable as the second differences, and must not be ignored when deriving the position of the BSCZ. We note that the signature in the ratios is supposed to be the direct signature of the BSCZ (not the residual of the signature once the HeII ionisation signal is removed) and the signature is often simpler than with other indicators for F-type stars (e.g. KIC6679371). 

Overall, if we assume that we are measuring the position of the BSCZ with all the three indicators, a solution -common to the three seismic indicators- is better retrieved with an expression that includes an additional term with twice the glitch period. The results of a fit of the $r_{010}$ performed with a theoretical expression including the extra term with twice the acoustic depth seems effective in finding the correct position of the BSCZ. The conclusions are more mitigated for the other indicators, especially the second differences. The main reason is the fact that the signal of the BSCZ for F-type stars is stronger in the ratios. Moreover, the helium signature being negligible in the ratios, the glitch signature of the BSCZ is easier to detect (i.e. there are fewer parameters to fit and less pollution in the signature). 

\section{Distributions of the acoustic depth of the surface convective zone for the intermediate stars (sample B)}\label{sec:sampleB}

In this section we analyse the two stars of sample B. There is again a degeneracy in the fits between values of $k>1$ and values of $k$ close to zero (similar to the standard case). To make the analysis of KIC10162436 easier, we apply a prior on the value of $\tau_\mathrm{cz}$ with $t_\mathrm{cz}/\mathcal{T}<0.5$ when fitting the ratios to follow the expectations of theoretical prediction of the position of the BSCZ (see Sect.~\ref{sect:comp_obs}).

\subsection{KIC1435467}

The results obtained for the synthetic frequencies of the reference star are presented in Fig.~\ref{fig:KIC143}. 

\textit{Standard fits:} The distributions of $\tau_\mathrm{cz}$ obtained for the synthetic data of Group 1 show one single peak for each seismic indicator that matches the position of the BSCZ inputted in the synthetic data for all indicators. This indicates that in presence of a pure sinusoidal signature, all indicators provide  the proper position of the BSCZ with a standard fit.
The dominant peak in the observed second differences distribution matches perfectly the one from the synthetic data (by construction). While the observed frequency-based  distribution is double-peaked, its highest amplitude peak does correspond to the one present in the synthetic distribution  at  the position of the BSCZ inputted. In the observed distribution of the $r_{010}$ ratios, the amplitude  of the peak at the position of the BSCZ inputted almost vanishes leaving a dominant high amplitude peak at a position completely off (about $1500$~s) that of the inputted BSCZ value.

\textit{Non-sinusoidal fits:} When the observed distributions are compared to the Group  2 synthetic distributions, not only the distribution from the second differences (by construction) but also the distributions of the other two indicators match perfectly the ones from the synthetic data. The three observed indicators give a coherent answer, with all three dominant peaks coinciding at the same $\tau_\mathrm{cz}$ position.

\paragraph{Conclusion}
We found that the standard theoretical glitch expression for the BSCZ does not provide a common solution for the three indicators. In contrast, when including an additional term in the theoretical glitch expression, we find a clear common solution between the three seismic indicators.

\subsection{KIC10162436}

The results obtained for KIC10162436 are presented in Fig.~\ref{fig:KIC1016}.

\textit{Standard fits:} Similarly to the case of KIC1435467, the Group 1  synthetic data provide clear distributions for the three seismic indicators with a dominant peak at the position of the BSCZ used to construct the synthetic data. However, they cannot reproduce the complex patterns of the observed distributions. Nevertheless, the dominant peak of the observed distributions of the second differences and the ratios coincides with the dominant one for the synthetic data at the position of the inputted position of the BSCZ in the synthetic data.

\textit{Non-sinusoidal fits:} For each indicator, the observed distribution is compared to the Group 2 synthetic distribution, both fitted with a non-sinusoidal formulation. One finds that the observed dominant peak in the $r_{010}$ coincides with the dominant one for the synthetic data at the position of the inputted position of the BSCZ in the synthetic data. For the two other indicators, the observed distributions are noisy, but in both cases, a peak - albeit with a small amplitude - is present that coincides with the dominant peak in the synthetic data at the position of the inputted position of the BSCZ. 

\paragraph{Conclusion} 
No clear common solution between the observed distributions of the three seismic indicators is obtained when using the standard theoretical glitch expression. The patterns in the three observed distributions are multi-modal and it is then difficult in such a case to derive the position of the BSCZ with confidence. When including an additional term in the theoretical expression, only the ratio-based distribution is single peaked at the right position.

\subsection{Interpretation of the results}

For this sample of low temperature F-type stars, clear conclusions are more difficult to draw. The measured position of the convective envelope with the additional term is sometimes deeper than with the standard fit (KIC10162436) and in both cases much deeper than the theoretical predictions (see Sect.~\ref{sect:comp_obs}). The analysis of the glitch signature of the low temperature F-type star seems more complicated than for the hottest ones. The additional term used for the analyse may also be too simple or not adapted in this case.

\section{Analysis of observed $r_{010}$ glitch signatures}\label{sect:comp_obs}

In Appendix~\ref{Apdx:fits} we show and discuss the results of the fits obtained from the observed data for three \textit{Kepler} stars (KIC1435467, KIC6679371, KIC10162436) when using the non-sinusoidal expression of the $r_{010}$ ratios. When comparing directly the signatures in the ratios $r_{010}$ with the associated fits, the $\chi^2$ values do not enable us to conclude in favour of one or the other (purely sinusoidal or not) analytical expressions of the glitch. This would require ratios with smaller error bars.

\begin{figure}
    \centering
    \includegraphics[scale=0.62]{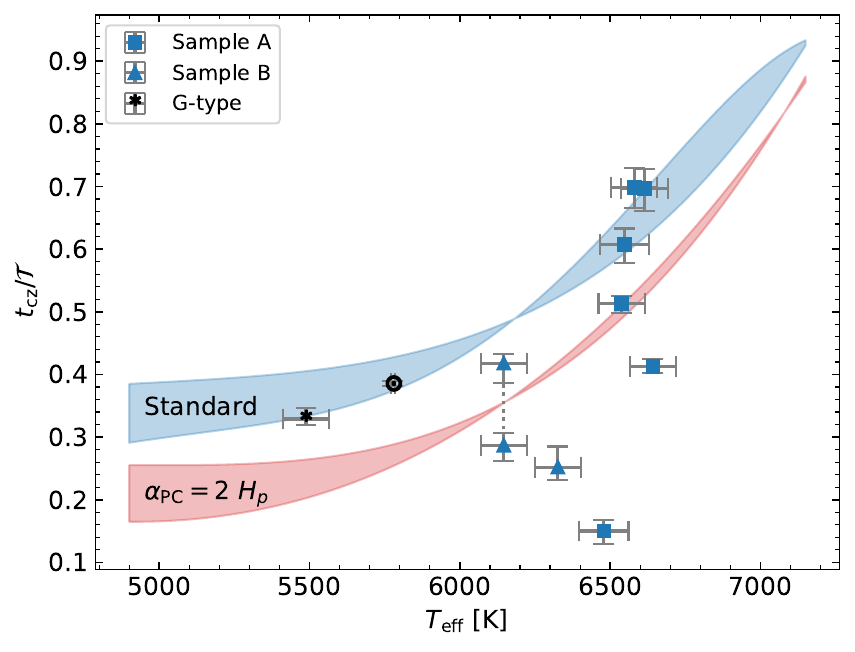}
    \includegraphics[scale=0.62]{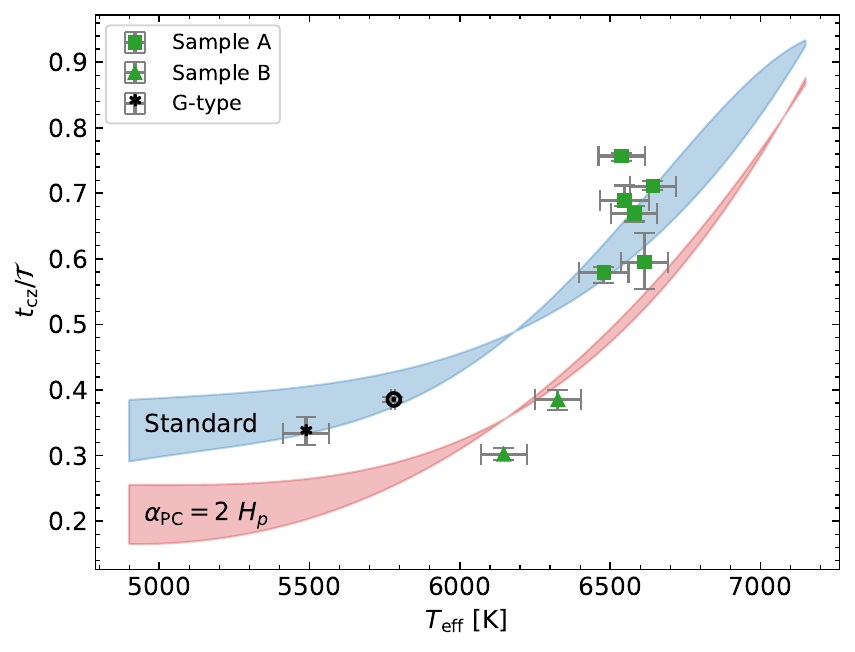}
   \caption{Comparison of the measured positions of the BSCZ in the $r_{010}$ ratios with predictions of standard models (blue area) and models including a penetrative convection of $2~H_p$ for the three \textit{Kepler} stars. The Sun is represented by the solar black symbol and Doris (KIC8006161) by the black star symbol. The models predictions are extended regions, rather than lines, because they included the position of the BSCZ from the zero age main sequence to the terminal age main sequence. The top panel presents the measurement with the standard fitting expression and the bottom panel presents the results with the non-sinusoidal fitting expression.}
   \label{fig:comp_mods}
\end{figure}

From the stellar model point of view, the position of the base of convective envelope is predicted to be around $0.47<t_\mathrm{cz}/\mathcal{T}<0.60$ for stars with $6100<T_\mathrm{eff}<6500$~K.
Figure~\ref{fig:comp_mods} shows the expected BSCZ position for standard models (blue area) and for models including a penetrative convection of $2~H_p$ (red area). The expected trend with effective temperature is that the shallower the BSCZ, the higher the $T_\mathrm{eff}$.

When the position of the convective envelope is measured in the $r_{010}$ ratios for the three samples (see Sect.~\ref{sec:samples}) with the standard signature expression (Eq.~\ref{eq:fit_r010_std}), the fitted acoustic depths are found much larger compared to those predicted by stellar models (see top panel of Fig.~\ref{fig:comp_mods} and Table~\ref{tab:2}). For these stars, the fitted acoustic depth is even larger than when assuming a $2~H_p$ overshoot penetration. Further, the trend with effective temperature is  found opposite to what is expected from stellar models. For KIC1435467, the glitch signatures seen in the second differences could be explained only by an extension of the PC region larger than $2~H_p$, which is very improbable. 

When the position of the convective envelope is measured with the non-sinusoidal expression, it is in better agreement with the predictions of stellar models and is not deeper than a $2~H_p$ PC region (see Table~\ref{tab:2}) except for KIC10162436. Moreover, the trend with effective temperature is in agreement with the stellar model predictions. The more striking example is KIC6679371 for which the position of the convective envelope is in perfect agreement with the predictions. We also see that for all the other stars of sample A, the position of the BSCZ measured with the non-sinusoidal fit is in good agreement with the model predictions. This illustrates the possible non-negligible non-sinusoidal contribution to the glitch signatures. 

\section{Discussion}\label{sect:discu}

\subsection{Pollution of the signature by the magnetic activity}

\begin{figure}
    \centering
    \includegraphics[scale=0.60]{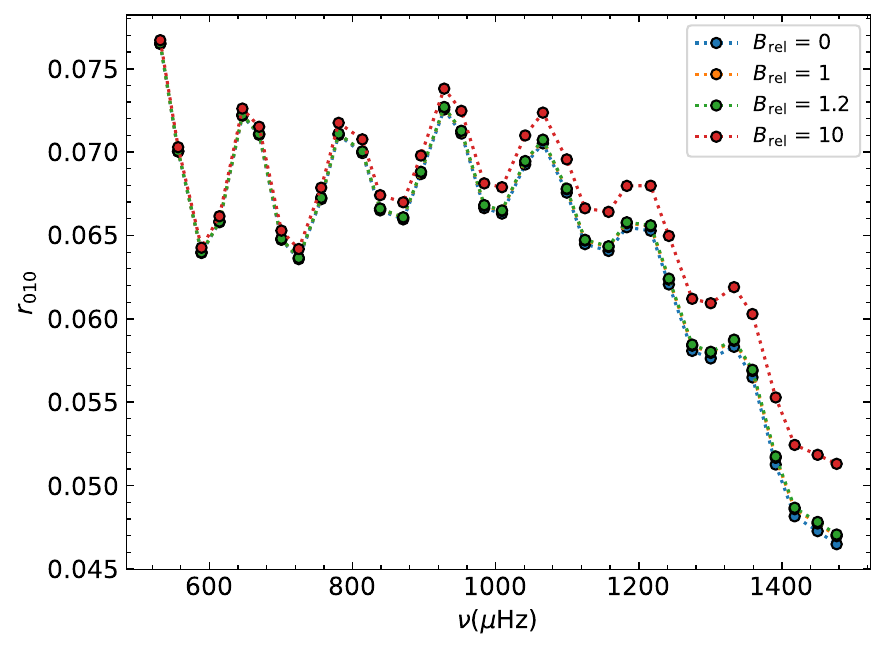}
   \caption{$r_{010}$ ratios according to the frequency for a $M=1.40$~M$_\odot$ model at $X_C=0.10$ for different values of relative-to-the-Sun magnetic field values $B_{\mathrm{rel}}$. We set $i=0$° and [$\lambda_\mathrm{min}$;$\lambda_\mathrm{max}$]=[$11$;$53$]. The blue, orange, and green points corresponding to $B_{\mathrm{rel}}= 0$, $1$, and $1.2$, respectively, are almost completely overlapping because the shift induced by the magnetic field is small.}
   \label{fig:mag}
\end{figure}

Despite the fact that $r_{010}$ are built with the purpose of not being sensitive to surface layers and the so-called surface effects, magnetic activity may have an impact if the associated near-surface perturbations are non-spherically symmetric. To assess the possible pollution of the glitch signature by magnetic activity, we modelled the frequency shift induced by the magnetic field following the work of \cite{thomas21}.

The B2 model of Paper I is used as reference ($M=1.40$~M$_\odot$, $X_C=0.10$, solar composition, penetrative convection of $\xi_\mathrm{PC}=2.0$ following \cite{zahn91} formalism -almost equivalent to $2.0~H_p$-). We apply values of the magnetic field of $1$, $1.2$, and $10$ times the solar value (the latter being an upper limit compared to the current knowledge of magnetic field in F-type stars: \citealt{anderson10,seach20}). We also adopt different values for the inclination angle $i$ of the star, and different latitudes $\lambda_\mathrm{min}$ and $\lambda_\mathrm{max}$ for the distribution of the activity at the surface of the star. The general effect is a shift of the ratios, with more impact at high frequencies, similarly to what is presented in \cite{thomas21} for the $r_{02}$ ratios (see Fig.~\ref{fig:mag}). However, the shift in the $r_{010}$ ratios is several orders of magnitude smaller than the uncertainties (about $0.0005$ for a magnetic field $1.2$ times the solar one, and $0.005$ for a magnetic field $10$ times the solar one, for $\nu=1400~\mu$Hz, $i=0$°, $\lambda_\mathrm{}$=[$11$;$53$]°, compared to uncertainties of about $0.015$ for KIC10162436), even when applying a magnetic field $10$ times larger than the solar one. Moreover, the shift of the ratios increases monotonically with the frequency, hence not disturbing the glitch signature from its original sinusoidal shape. From these tests, we conclude that magnetic activity (at least as we modelled it) has barely any impact on the $r_{010}$ ratios and on the measurement of the position of the BSCZ.

\subsection{Amplitude of the additional term}

\begin{figure}
    \centering
    \includegraphics[scale=0.60]{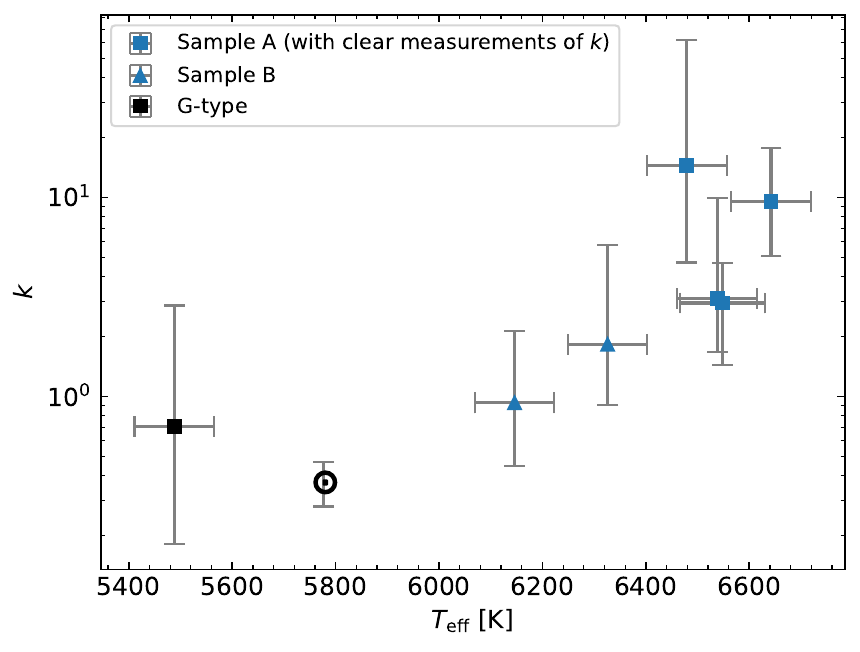}
   \caption{Values of $k$ according to the effective temperature obtained for the three samples with the fit of the ratios. For sample A, KIC2837475 and KIC12317678 are not shown because $k$ cannot be measured from the ratios (see Sect.~\ref{sec:KIC283}).}
   \label{fig:k}
\end{figure}

The relative amplitude of the additional term compared to the standard term is given by the coefficient $k$. We show its values according to the effective temperature of the stars in Fig.~\ref{fig:k}. We also add the $k$ parameters found when fitting the signature of the G-type stars KIC8006161 (Doris) and the Sun for comparison. It seems that both parameters are correlated, with an increase of $k$ with the temperature. This could indicate a larger contribution of the additional term when the convective envelope becomes shallower. The additional term to the glitch expression with twice the acoustic depth of the standard glitch expression can have several origins. We briefly discuss this topic below.

\subsection{Origin of the second order term in the glitch signature}

\noindent\textit{Second order term in the glitch signature:}\\
An asymptotic approach shows that the presence of a discontinuity in the structure quantities introduces a frequency correction that at first order in an asymptotic expansion, takes the form of a sinusoidal variation in the frequency of the p modes as a function of the radial order with a period and the acoustic depth $\tau$ \citep{gough90}. Starting from the second order formalism of \cite{tassoul80}, \cite{provost93} have extended the asymptotic calculation in presence of discontinuity of the sound speed to second order and have shown that the second order correction to the frequency is also a sinusoidal signal but with twice the frequency of the first order correction term. \cite{provost93} have assumed a discontinuity in the speed of sound itself. Here, if we attribute the glitch observed for the F-type stars to the discontinuity at the base of the convective zone in presence of  convective penetration, the transition between the radiative temperature gradient and the adiabatic convective temperature gradient is discontinuous. In such a case, the sound speed remains continuous at the transition but its derivatives are discontinuous. Preliminary calculations assuming a discontinuity in the sound speed first and second derivatives provide a resonant condition that leads to an $\arctan$ behaviour for the frequency correction with an argument in $O(1/\sigma)$ where $\sigma$ is the scaled frequency. As $1/\sigma$ is small for asymptotic p-mode frequencies, we find that the second-order frequency correction (and that of the ratio $r_{010}$) roughly behave as a sinusoid with a period $2\tau$ with an amplitude proportional to the difference in the sound speed derivatives on either side of the discontinuity. However, computing the amplitude of the glitch for a stellar model of a F-type star shows that the amplitude remains much too small compared with the observations (the details will be published elsewhere). It is possible that assuming a pure discontinuity at a single radius within the star is not enough and that a rapid variation over a larger region is more appropriate.\\

\noindent\textit{Large glitch:} The sinusoidal expressions introduced in Sect.~\ref{sect:theory} to represent the glitch signature on the pulsation frequencies were derived based on a variational analysis following \cite{monteiro94}. While in earlier works this has been found to be adequate to describe the signature of acoustic glitches, it is possible that in F-type stars a non-variational analysis is required to fully account for the glitch impact on the frequencies, as seems to be the case for buoyancy glitches. The perturbations to the frequencies of gravity and mixed modes caused by buoyancy glitches have been derived without recourse to a variational analysis by \cite{cunha15,cunha19b,cunha24}. Briefly, the authors derived the eigenvalue condition for a propagation cavity containing a structural glitch by matching the asymptotic eigenfunctions and their derivatives across the glitch. The glitch signature was found to be expressed as an arccot function, approaching the sinusoidal solution only when the glitch amplitude is sufficiently small. Such non-sinusoidal signatures, observed, for example in red-giant stars~\citep{vrard22}, can be expressed as a sum of a fundamental sinusoidal function and respective harmonics. In the specific case of buoyancy-glitch signature, the term with twice the acoustic depth in the frequency perturbations is expected to have a smaller amplitude than the base term, unlike what is seen in the data for the F stars. A similar derivation for the specific case of the acoustic glitch associated with the BSCZ is thus required and will be considered in future work.  \\

\noindent\textit{Non-linearity in the oscillations:}
The second term in the glitch signature might also be related to a non-linearity in the oscillations themselves. Indeed, the sinusoidal shape traditionally assumed for the glitch signature stems directly from the asymptotic expansion of the mode eigenfunctions. In broad outline, while the usual glitch signature involves the square of the displacement eigenfunction, if the amplitude of the modes is large enough, the added non-linear terms would involve the displacement eigenfunctions to the fourth power. In other words, the usual sinusoidal term with period $4\pi\tau_\mathrm{cz}$ would be supplemented with other sinusoidal terms with period $8\pi\tau_\mathrm{cz}$, which corresponds exactly to the additional term detected in the observations (see Eq.~\ref{eq:NL}).

To assess whether this sort of non-linear behaviour can explain the observed second term in the glitch signature, we performed a weakly non-linear expansion of the equation of hydrodynamics. Expanding the solution of each order of the expansion successively on the basis of the $p$-mode eigenfunctions, we eventually extracted a variational principle for the frequencies, and from it an expression for the glitch signature (i.e. the frequency shift $\delta\omega$ entailed by a perturbation in the sound speed profile in the form of a Dirac function). The details of the calculation will be the subject of a subsequent study. All calculations being carried out, we eventually obtain an expression for $\delta\omega$ of the form
\begin{multline}
    \delta\omega = A(\omega) + B(\omega) \cos 2 \theta_\mathrm{cz} + C(\omega) \sin 2 \theta_\mathrm{cz} \\
    + D(\omega) \cos 4 \theta_\mathrm{cz} + E(\omega) \sin 4 \theta_\mathrm{cz}~,
\end{multline}

\noindent where $\theta_\mathrm{cz}$ is a phase that goes as $\omega \tau_\mathrm{cz}$, and the coefficients $A$ through $E$ depend on i) the conditions at the base of the convective zone (sound speed and density), ii) the amplitude of the acoustic modes, and iii) the frequency $\omega$ itself. This expression is equivalent to Eq.~\ref{eq:NL}. In the linear case, i.e. when the amplitude of the modes tend to $0$, only the $2\tau_\mathrm{cz}$ period remains, and the expression reduces to the usual glitch signature \citep{monteiro94,roxburgh94}.

\begin{figure}
    \centering
    \includegraphics[width=\linewidth]{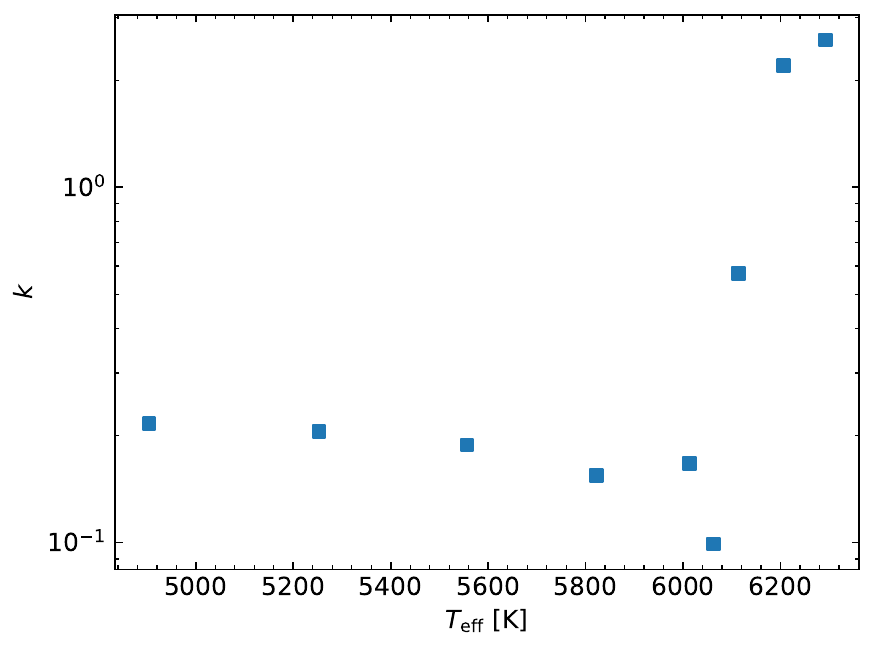}
    \caption{Values of $k$ according to $T_\mathrm{eff}$ for the stellar models considered in the text for mode amplitudes eight orders of magnitude higher than observed. A clear change of regime can be found between G- and F-type stars, which is in line with the observational results obtained in this study.}
    \label{fig:non-linear-k}
\end{figure}

We applied these expressions, and extracted a value for $k$ (see Eq.~\ref{eq:NL}), for the same set of stellar models that we used for Fig.~\ref{fig:predi} and the blue area of Fig.~\ref{fig:comp_mods}), spanning effective temperatures between $4,700$ K and $6,300$ K, obtained with the \texttt{Cesam2k20} stellar evolution code. Adopting mode amplitudes of $1~m/s$, we found that $k$ remained much smaller than $1$ for all stars, meaning that we were not able to connect the observed second term in the glitch signature to a non-linear behaviour of the oscillations. However, we show in Fig.~\ref{fig:non-linear-k} the dependence of $k$ with $T_\mathrm{eff}$ when we increase the amplitude $A_\mathrm{obs}$ by a factor $10^8$. It clearly shows a change of behaviour between G-type and F-type stars: for the former, the ratio $k$ remains smaller than $1$ and fairly independent of $T_\mathrm{eff}$; for the latter, we find an increase of $k$ with $T_\mathrm{eff}$, with values that exceed $1$ for the hotter stars. This is in line with the results presented in this study, and in particular with the plot shown in Fig.~\ref{fig:k}, with a change of regime for $T_\mathrm{eff} \sim 6,000$ K. Even though the mode amplitude needed for this is orders of magnitude too large compared to observed mode amplitudes in these stars, this result might indicate that the origin of the extra term in the glitch signature is to be looked for beyond a purely linear formalism.

\section{Conclusion}\label{sect:conclu}

We analysed the glitch signatures in nine \textit{Kepler} stars and the Sun using the frequencies, second differences, and $r_{010}$ ratios. We focused on the measurements of the position of the BSCZ and showed that for the F-type stars of the sample, the signatures point to different values for the measurement of the acoustic depth of the BSCZ when using the usual fitting expressions for the three seismic indicators. In contrast, this was not found for G-type stars, for which all seismic indicators agree. We also found that the posterior distributions obtained for the acoustic depth of the BSCZ are more difficult to interpret in both the frequencies and the second differences (multi-peaks distributions) than in the $r_{010}$ ratios (mostly single-peak distributions). Possible explanations are that we observed glitch signatures with different origins in the three seismic indicators, that the data quality is not high enough to extract useful information, or that the fitting expressions are not suitable for F-type stars.

Assuming that only the bottom of the convective envelope can be responsible for such signatures, the measurements provided by the standard fitting expressions indicated very unexpected depths for convective envelopes of F-type stars. Considering that the above assumption is still valid, we then explored the possibility of including an additional term in the fitting expression with twice the acoustic depth of the standard term, thus accounting for a possible non-linear contribution. This approach was purely theoretical since the standard fitting expressions are  already able to reproduce  the data for F-type stars (due to the large uncertainties on the individual frequencies). We show that in most of the cases when the second differences and the ratios provide different values for the acoustic depth of the BSCZ, the value obtained from the ratios corresponds to twice that obtained with the second differences. We also find that such an expression provides measurements of the acoustic depths of the BSCZ that are in better agreement with the prediction of stellar evolution models. We note that the standard fitting expressions often point to the position of the BSCZ as being deeper than $2~H_p$ from the predictions of stellar models, which is far from the theoretical predictions. Moreover, we showed that applying this extra term to the fit of the signatures of G-type stars does not affect the measurement of the BSCZ at all.

The physical origin of the extra term in the theoretical glitch formulation is not yet clear and will require additional theoretic investigations. The most promising explanation comes from the theoretical modelling of the glitch that is currently treated as a perturbation, but this may not be the case for F-type stars. Nevertheless, understanding these signatures in F-type stars will bring valuable information about the physics of these stars and improve their modelling. The future asteroseismic space missions (e.g. PLATO) will also strongly help in this regard.

\begin{acknowledgements}
We gratefully thank our anonymous referee whose comprehensive readings and remarks helped to improve the content of the manuscript. This work was supported by CNES, focused on PLATO. This work was supported by FCT/MCTES through the research grants UIDB/04434/2020, UIDP/04434/2020 2022.06962.PTDC., 2022.03993.PTDC, and DOI 10.54499/2022.03993.PTDC. MD thanks Julien Morin for fruitful discussions about stellar magnetic fields.
\end{acknowledgements}  

\bibliographystyle{aa} 
\bibliography{main.bib} 
 
\newpage

\begin{appendix}
\onecolumn
\section{Distributions of the glitch signature in the frequencies, second differences, and $r_{010}$ ratios for sample A }

\begin{figure}[h!]
    \centering
    \includegraphics[scale=0.55]{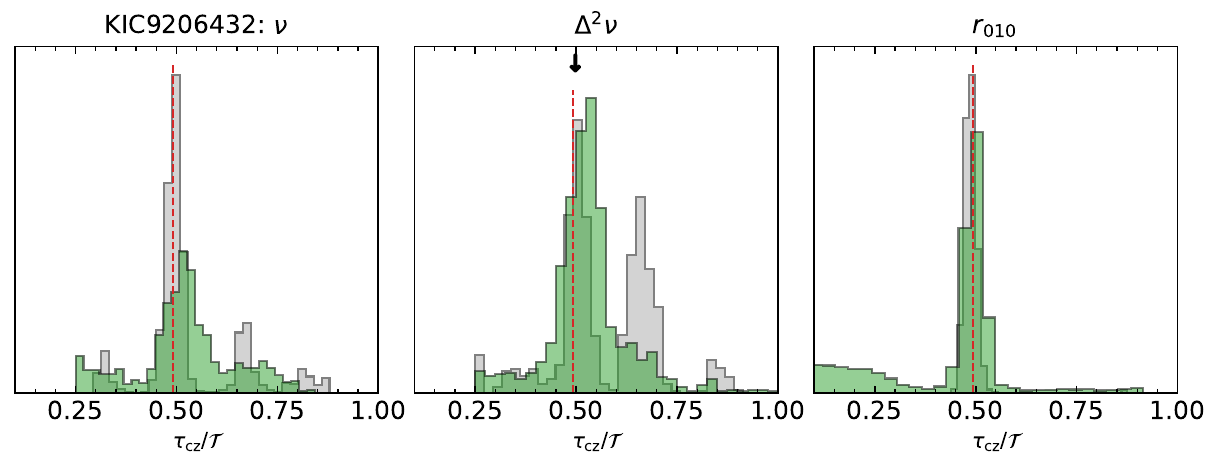}
    \includegraphics[scale=0.55]{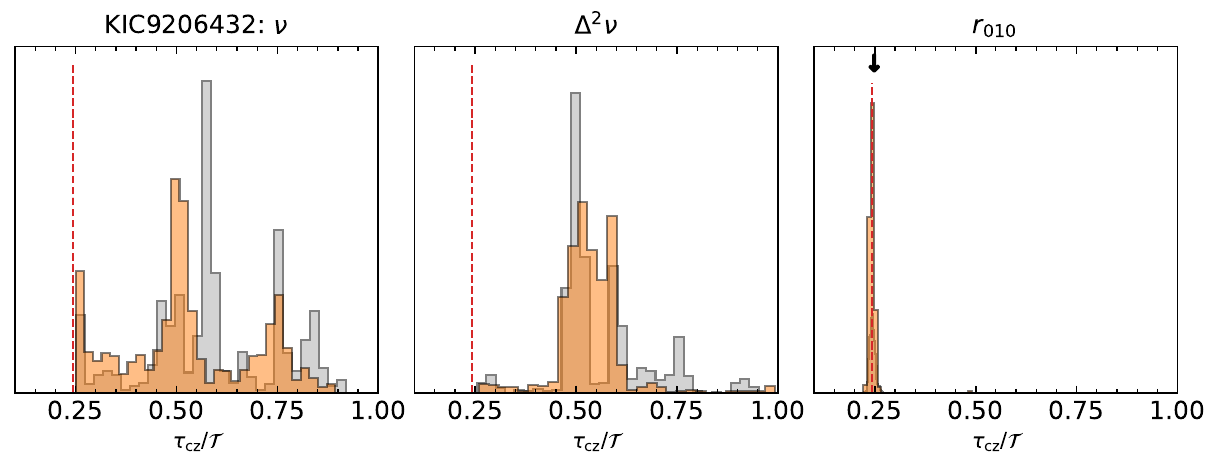}
   \caption{Same as Fig.~\ref{fig:KIC667} but for KIC9206432.}
   \label{fig:KIC920}
\end{figure}

\begin{figure}[h!]
    \centering
    \includegraphics[scale=0.55]{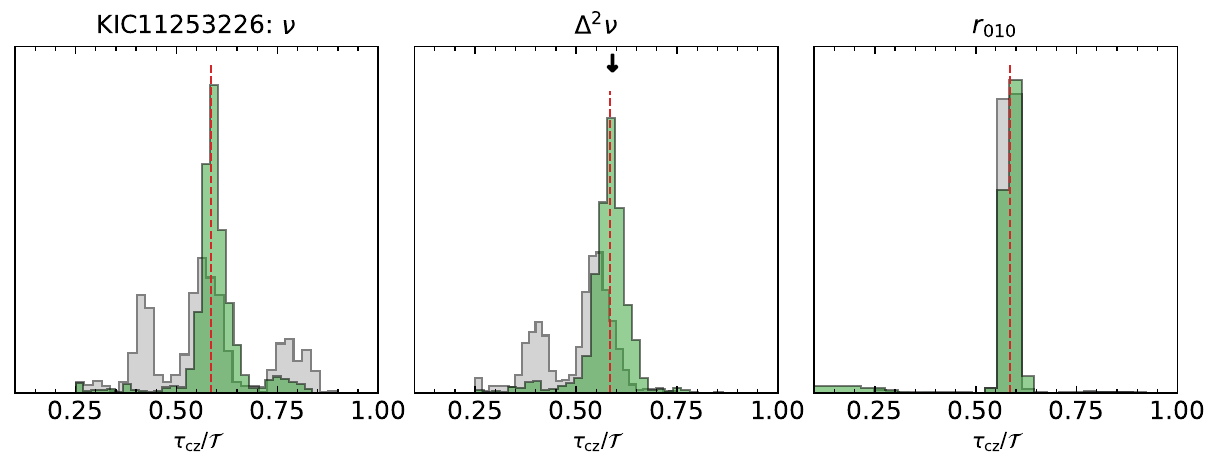}
    \includegraphics[scale=0.55]{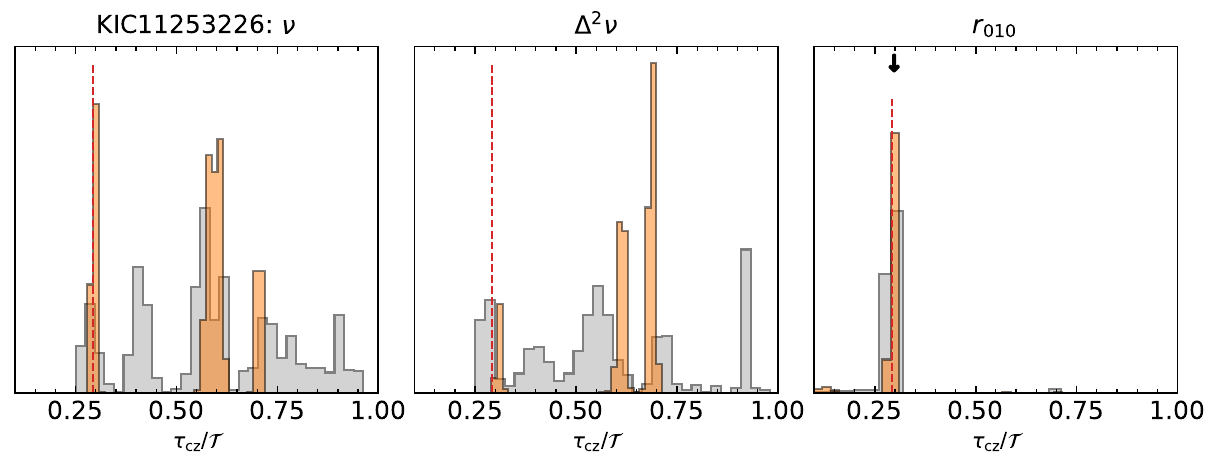}
   \caption{Same as Fig.~\ref{fig:KIC667} but for KIC11253226.}
   \label{fig:KIC112}
\end{figure}
\begin{figure*} 
    \centering
    \includegraphics[scale=0.55]{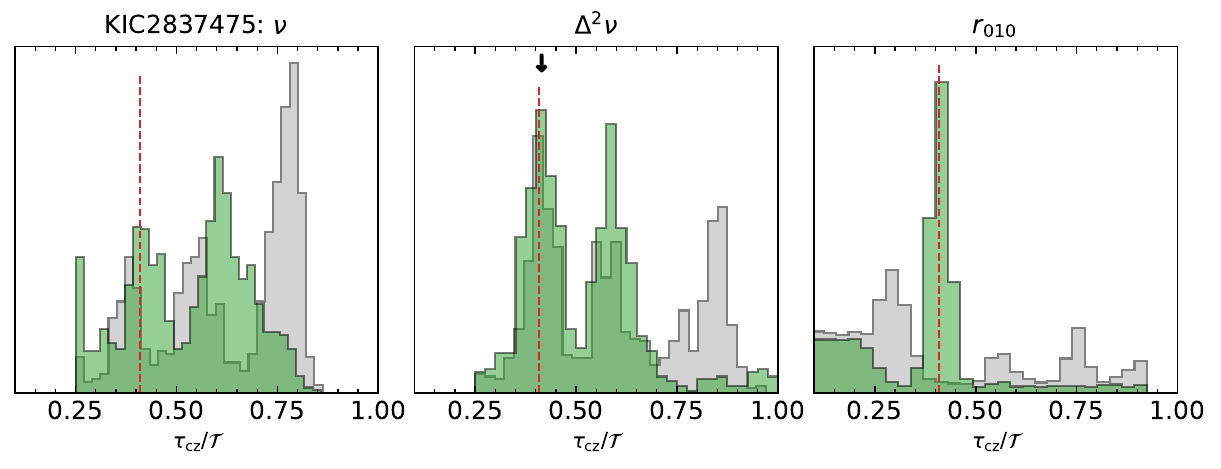}
    \includegraphics[scale=0.55]{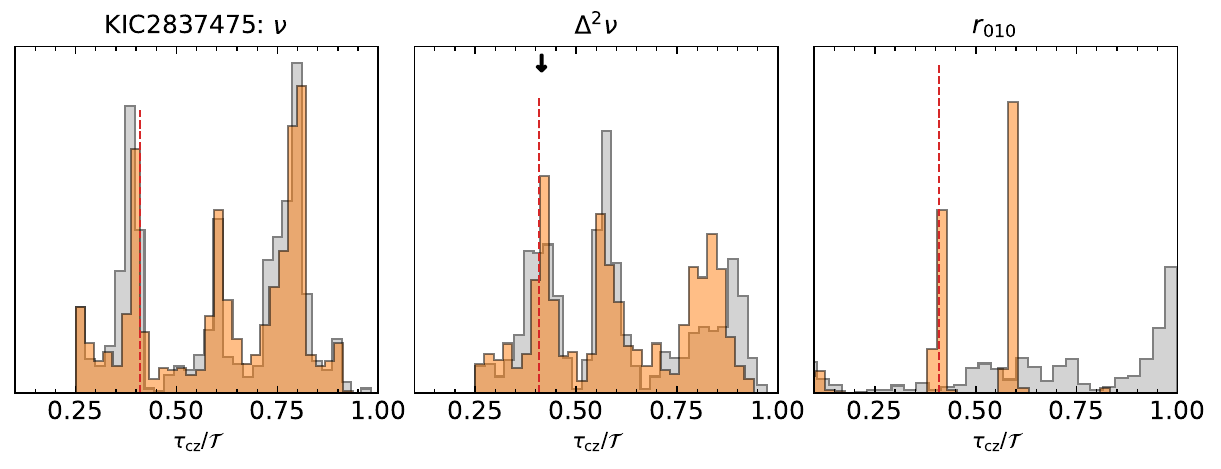}
    \put(-50,100){$k=4.0$}\\
    \includegraphics[scale=0.55]{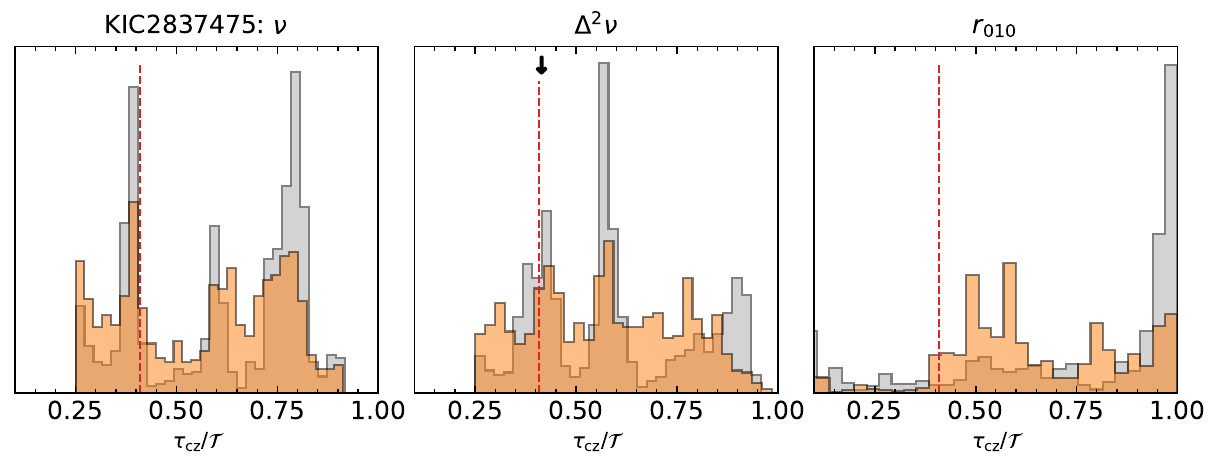}
    \put(-50,100){$k=1.5$}
   \caption{Same as Fig.~\ref{fig:KIC667} but for KIC2837475. The middle and bottom panel correspond to different value of $k$.}
   \label{fig:KIC283}
\end{figure*}
\begin{figure*} 
    \centering
    \includegraphics[scale=0.55]{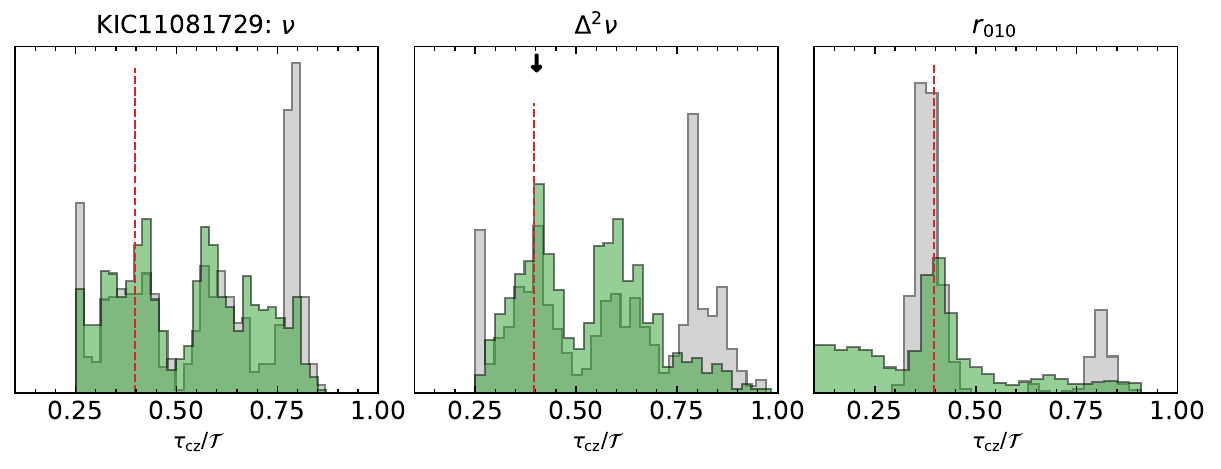}
    \includegraphics[scale=0.55]{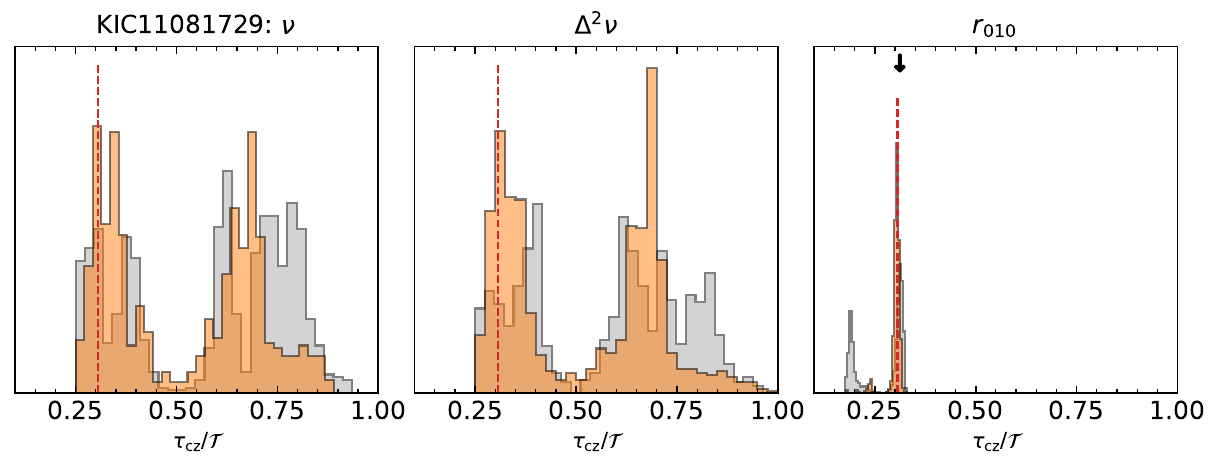}
   \caption{Same as Fig.~\ref{fig:KIC667} but for KIC11081729.}
   \label{fig:KIC110}
\end{figure*}

\begin{figure*} 
    \centering
    \includegraphics[scale=0.53]{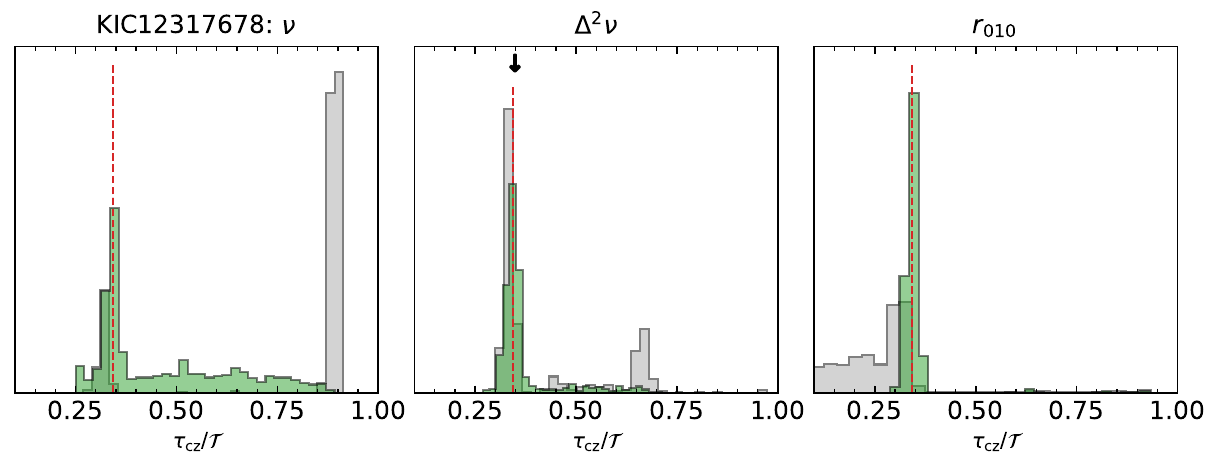}
    \includegraphics[scale=0.53]{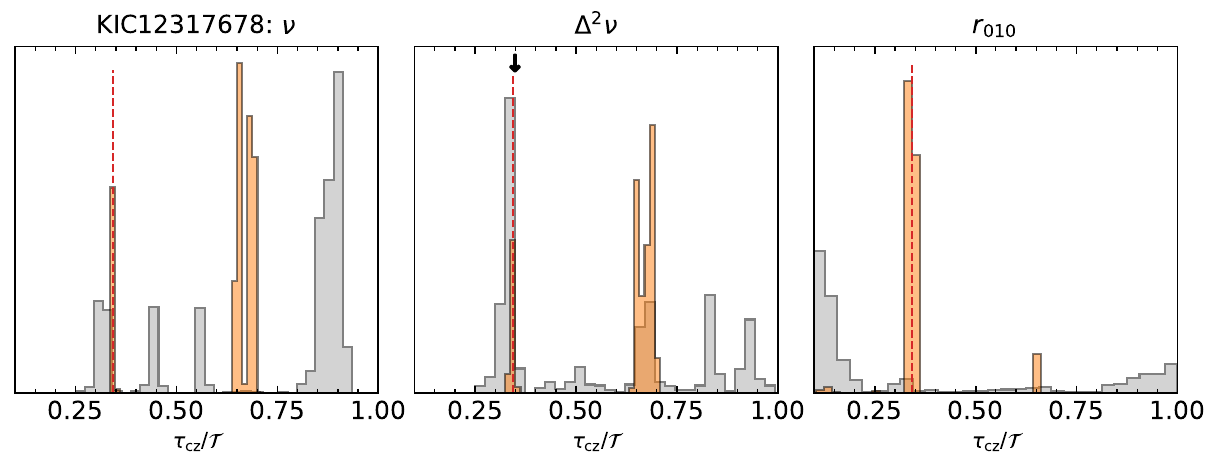}
    \put(-55,95){$A_2=0.25$}\\
    \includegraphics[scale=0.53]{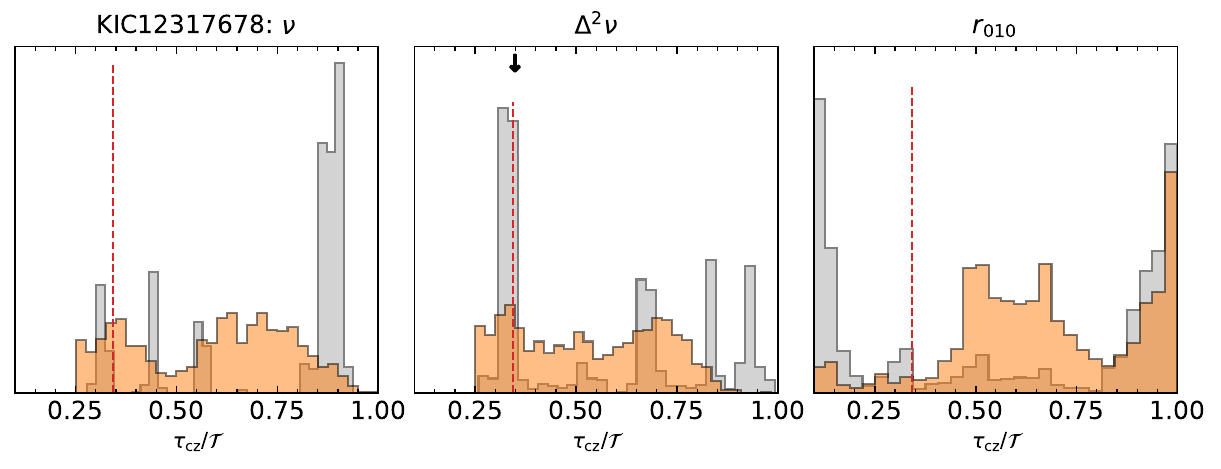}
    \put(-60,95){$A_2=0.025$}
   \caption{Same as Fig.~\ref{fig:KIC667} but for KIC12317678. The middle and bottom panel correspond to different values of $A_2$.}
   \label{fig:KIC123}
\end{figure*}

\FloatBarrier
\section{Distributions of the glitch signature in the frequencies, second differences, and $r_{010}$ ratios for sample B }

\begin{figure}[h!]
    \centering
    \includegraphics[scale=0.53]{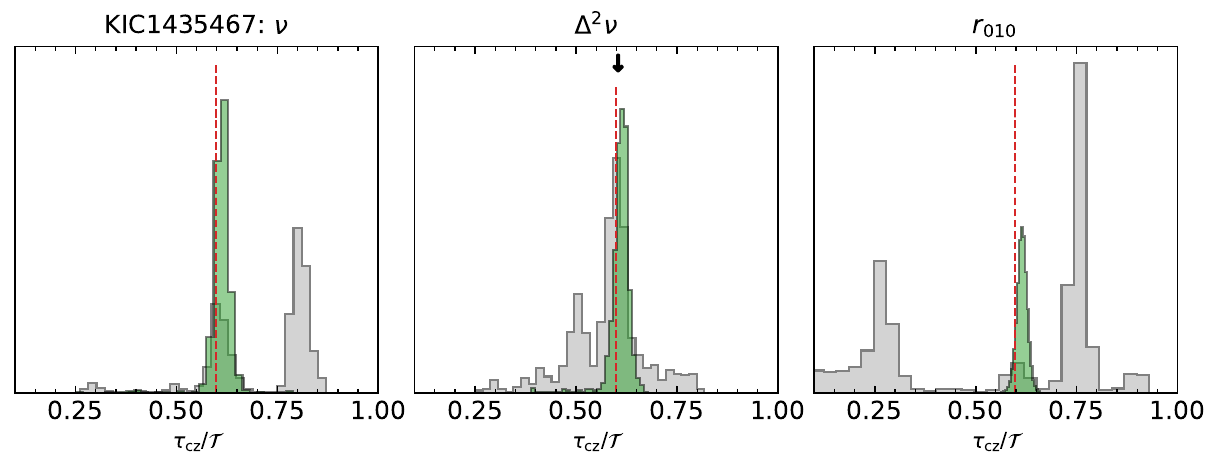}
    \includegraphics[scale=0.53]{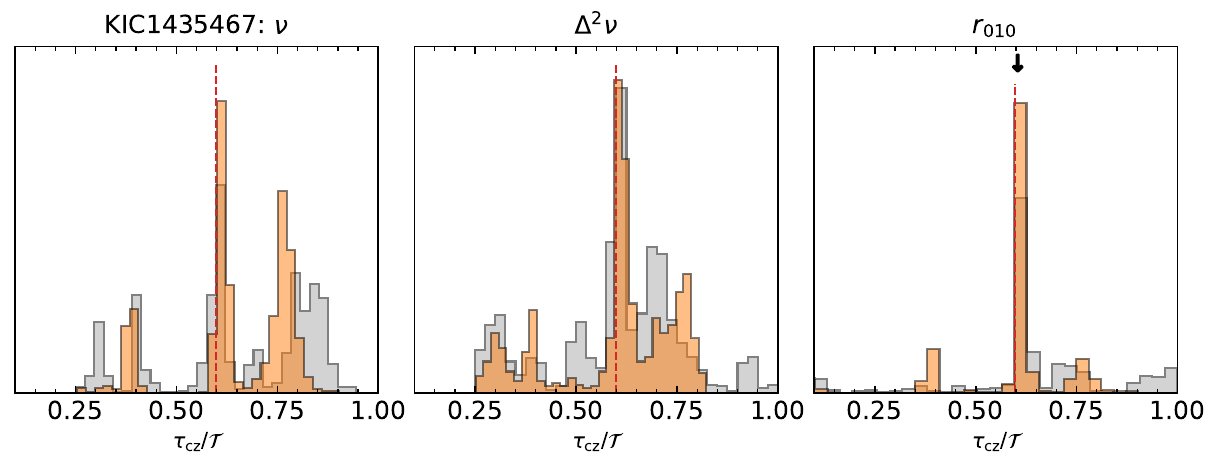}
   \caption{Same as Fig.~\ref{fig:KIC667} but for KIC1435467.}
   \label{fig:KIC143}
\end{figure}

\begin{figure}[h!]
    \centering
    \includegraphics[scale=0.55]{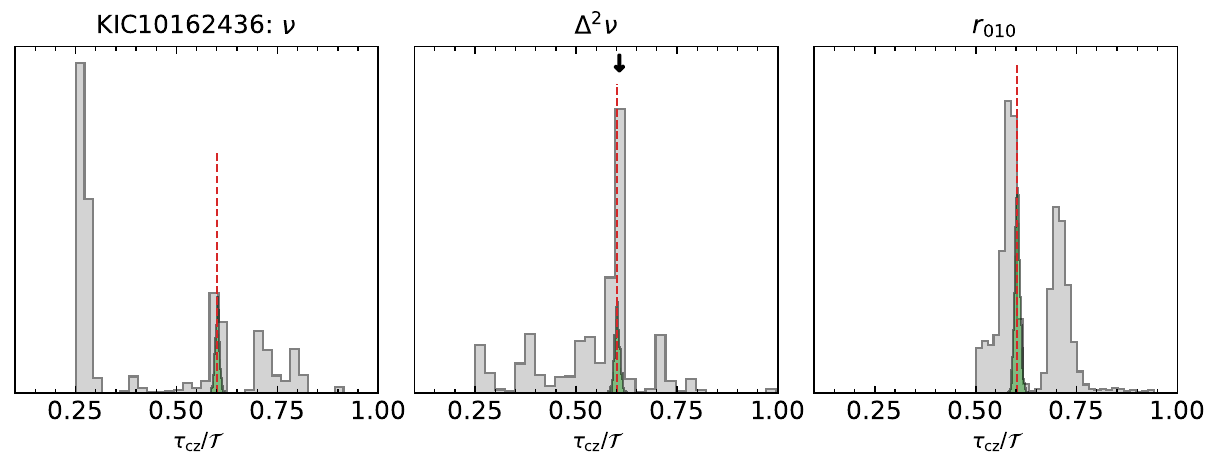}
    \includegraphics[scale=0.55]{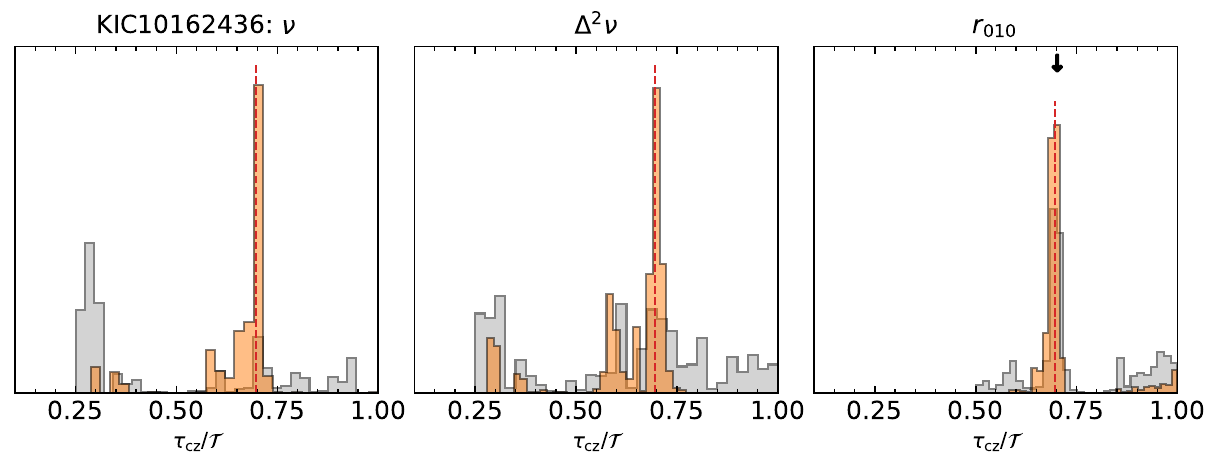}
   \caption{Same as Fig.~\ref{fig:KIC667} but for KIC10163436.}
   \label{fig:KIC1016}
\end{figure}
\FloatBarrier
\section{Results of the $r_{010}$ fits}\label{Apdx:fits}
 
\begin{figure*}
    \centering
    \includegraphics[scale=0.55]{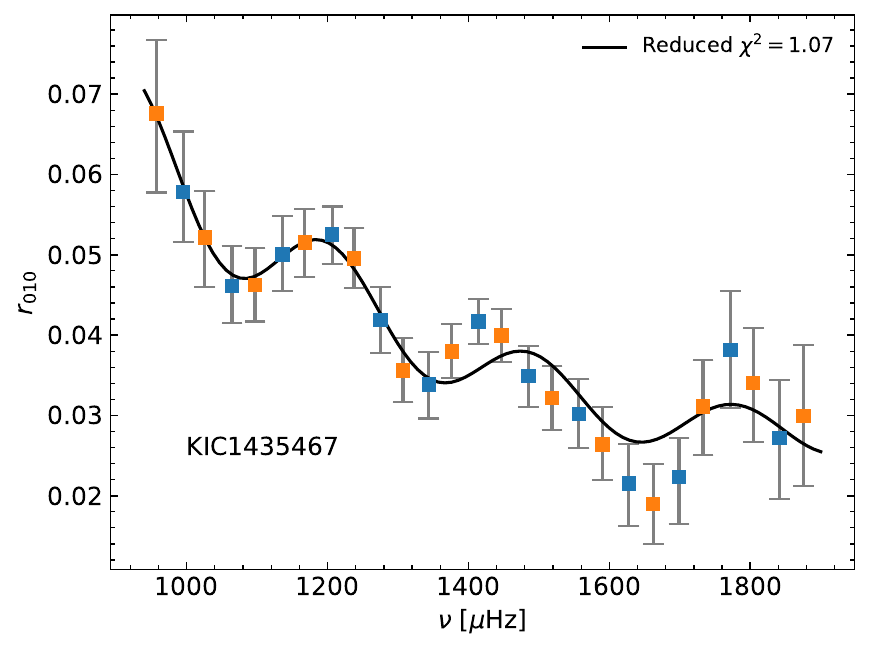}
    \includegraphics[scale=0.55]{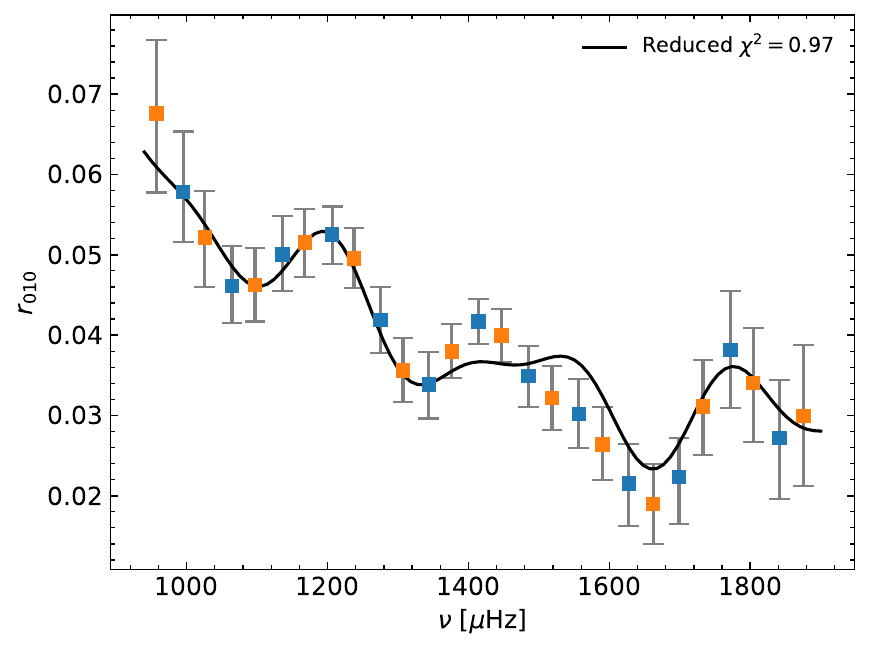}
   \caption{$r_{010}$ ratios according to the frequency for the star KIC1435467. The left panel shows fit when the standard expression is taken into account. The right panel show the fit when the non-sinusoidal expression is used. The blue and orange symbols represent the seismic indicators for the $l=0, 1$ modes, respectively. The black curves represent the fit of the seismic indicators.}
   \label{fig:fit143}
\end{figure*}

\begin{figure*}
    \centering
    \includegraphics[scale=0.6]{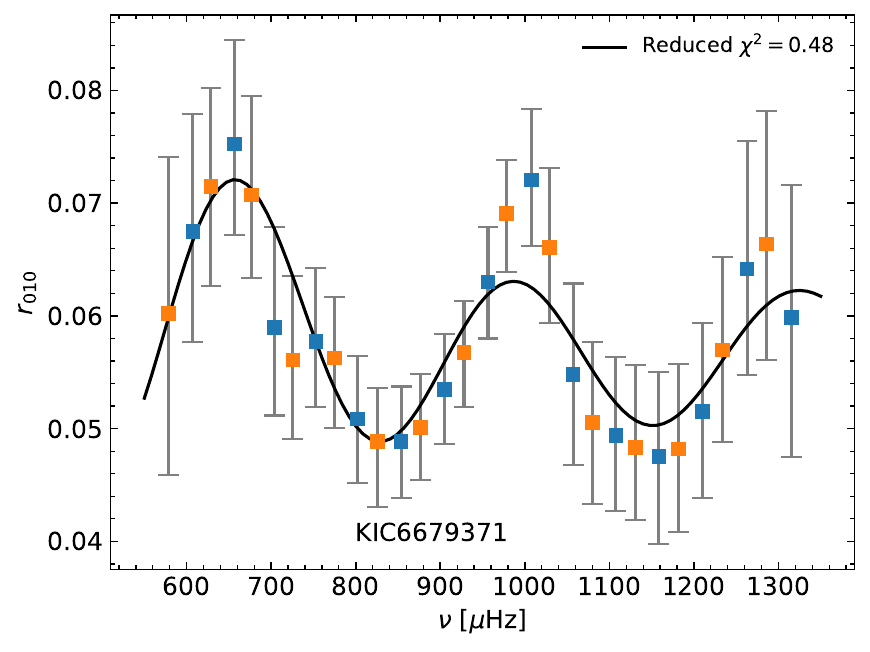}
    \includegraphics[scale=0.6]{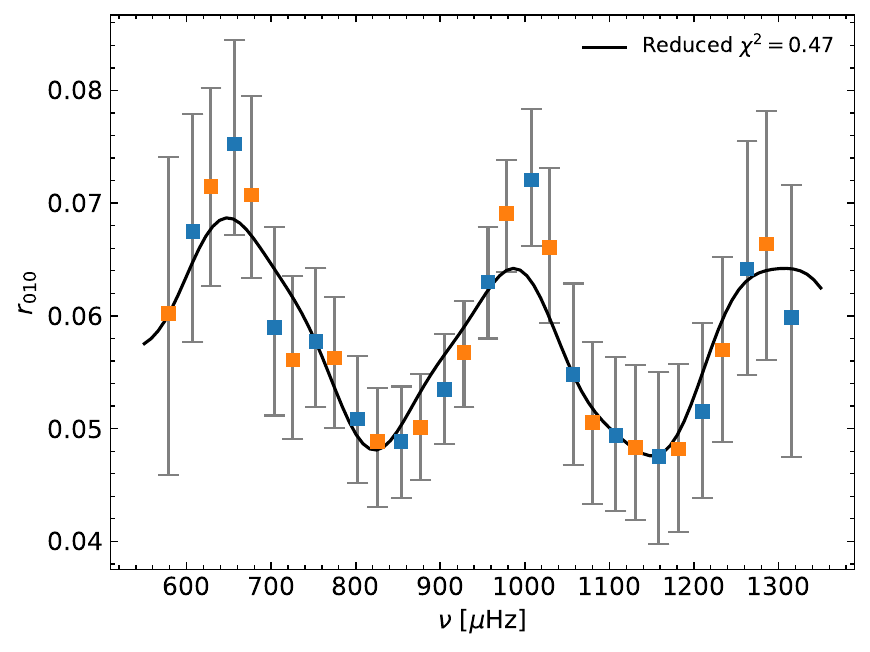}
   \caption{Same as Fig.~\ref{fig:fit143} but for the star KIC6679371.}
   \label{fig:fit667}
\end{figure*}

\begin{figure*}
    \centering
    \includegraphics[scale=0.6]{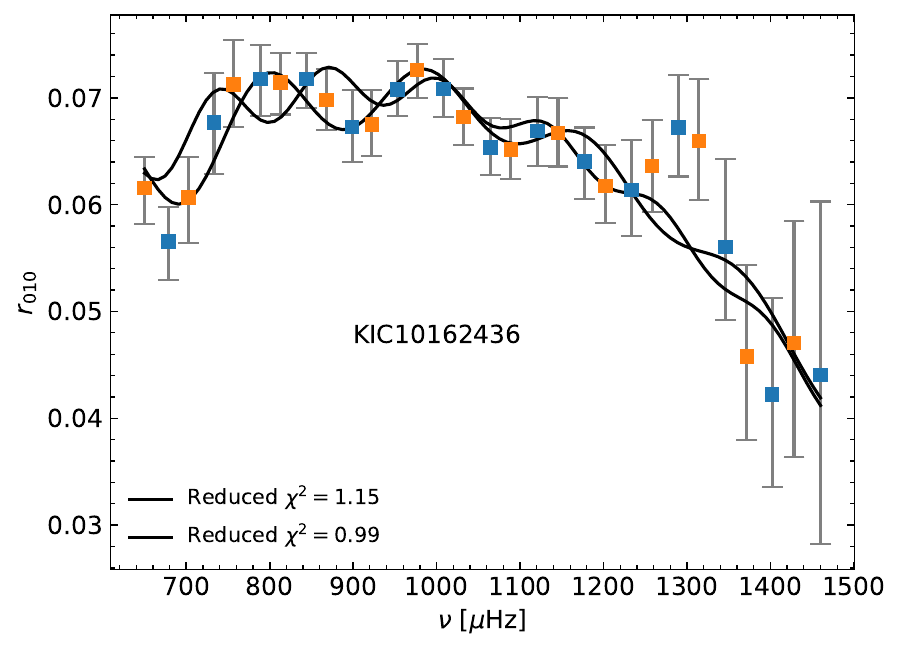}
    \includegraphics[scale=0.6]{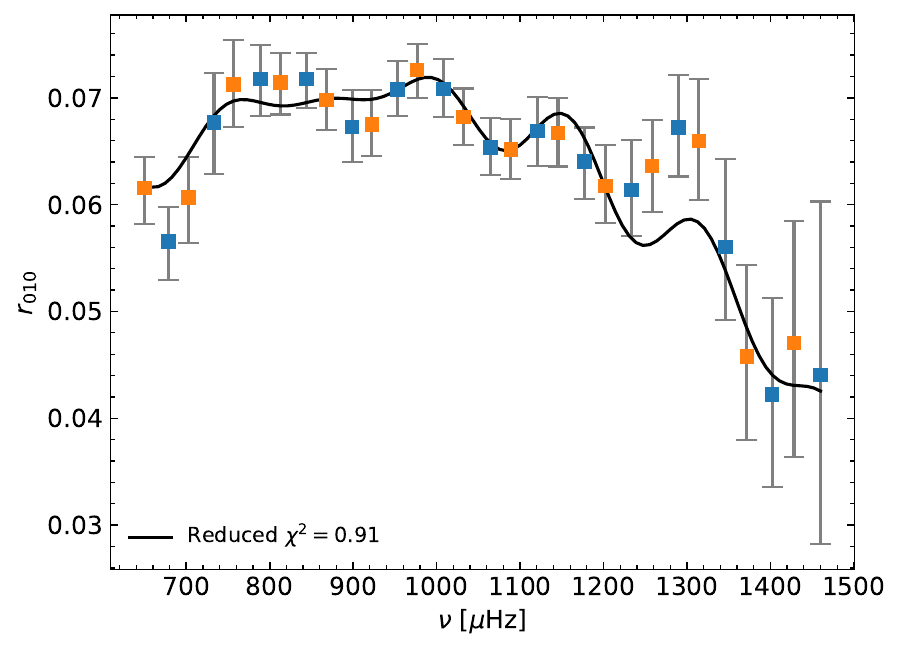}
    \includegraphics[scale=0.6]{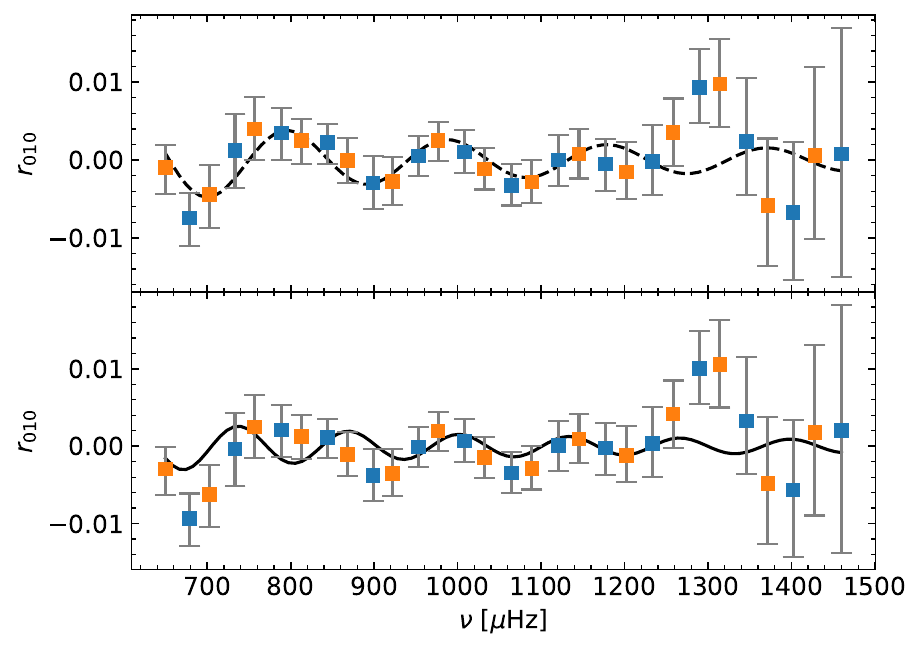}
   \caption{Same as Fig.~\ref{fig:fit143} but for the star KIC10162436 for the top panels. The left panel shows the two solutions found when the standard expression is taken into account. Both solutions are separated in the bottom panel, where the smooth component has been removed from the signal. }
   \label{fig:fit1016}
\end{figure*}

In this appendix we show the fit of the ratios for three stars. The reduced $\chi^2$ is computed with 
\begin{equation}
    \chi^2_R=\frac{\left(r_{010,\mathrm{obs}}-r_{010,\mathrm{fit}}\right)^T\mathcal{C}^{-1}\left(r_{010,\mathrm{obs}}-r_{010,\mathrm{fit}}\right)}{N},
\end{equation} 
\noindent where $\mathcal{C}$ is the covariance matrix, $r_{010,\mathrm{obs}}$ are the observed ratios, $r_{010,\mathrm{fit}}$ are the ratios obtained from the fit, and $N$ is the degree of freedom (number of data point minus the number of free parameters, that is seven for the sinusoidal fit and eight for the non-sinusoidal). For KIC1435467, the non-sinusoidal expression gives a $\chi^2$ only slightly better (Fig.~\ref{fig:fit143}). For this star, we find $k=2.02^{+2.44}_{-1.10}$ with the non-sinusoidal fit. For KIC6679371, only one periodic signal seems to dominate the signature, so both expressions seem to perform equivalently with a slightly better $\chi^2$ again for the non-sinusoidal fit (Fig.~\ref{fig:fit667}). In this case, we find $k=14.48^{+47.23}_{-9.77}$. The results of the fit are presented in Fig.~\ref{fig:fit1016} for KIC10162436. First, for the standard fitting expression, we find a double peak solution and the $\chi^2$ a preference for the low $t_\mathrm{cz}$ solution. The non-sinusoidal fit has only one solution and also favour the same solution. Even if the fitting with the non-sinusoidal expression seems better by eye, the reduced $\chi^2$ is slightly larger. For the non-sinusoidal fit we find $k=0.93^{+1.20}_{-0.48}$. At the end, we cannot really favour one or the other fitting expression directly from the data for the three stars. For that, smaller error bars on the ratios would be necessary. 

\nolinenumbers

\end{appendix}
\end{document}